\documentclass[%
 aip, 
 amsmath,amssymb,
reprint,%
]{revtex4-1}

\usepackage{graphicx}
\usepackage{dcolumn}
\usepackage{bm}

\usepackage[utf8]{inputenc}
\usepackage[T1]{fontenc}
\usepackage{mathptmx}
\usepackage{etoolbox}

\usepackage{algorithm}
\usepackage{algpseudocode}
\usepackage{xcolor}

\makeatletter
\def\@email#1#2{%
 \endgroup
 \patchcmd{\titleblock@produce}
  {\frontmatter@RRAPformat}
  {\frontmatter@RRAPformat{\produce@RRAP{*#1\href{mailto:#2}{#2}}}\frontmatter@RRAPformat}
  {}{}
}%
\makeatother
\begin{document}


\title{Deep reinforcement learning for large-eddy simulation modeling in wall-bounded turbulence}
\author{Junhyuk Kim}
\affiliation{%
	Department of Mechanical Engineering, Yonsei University, Seoul 03722, Republic of Korea
}%
\author{Hyojin Kim}
\affiliation{%
	Department of Mechanical Engineering, Yonsei University, Seoul 03722, Republic of Korea
}%
\author{Jiyeon Kim}
\affiliation{%
	School of Mathematics and Computing, Yonsei University, Seoul 03722, Republic of Korea
}%
\author{Changhoon Lee} \email{clee@yonsei.ac.kr}
\affiliation{%
	Department of Mechanical Engineering, Yonsei University, Seoul 03722, Republic of Korea
}%
\affiliation{%
	School of Mathematics and Computing, Yonsei University, Seoul 03722, Republic of Korea
}%

\date{\today}

\begin{abstract}
The development of a reliable subgrid-scale (SGS) model for large-eddy simulation (LES) is of great importance for many scientific and engineering applications. Recently, deep learning approaches have been tested for this purpose using high-fidelity data such as direct numerical simulation (DNS) in a supervised learning process. However, such data are generally not available in practice. Deep reinforcement learning (DRL) using only limited target statistics can be an alternative algorithm in which the training and testing of the model are conducted in the same LES environment. The DRL of turbulence modeling remains challenging owing to its chaotic nature, high dimensionality of the action space, and large computational cost. In the present study, we propose a physics-constrained DRL framework that can develop a deep neural network (DNN)-based SGS model for the LES of turbulent channel flow. The DRL models that produce the SGS stress were trained based on the local gradient of the filtered velocities. The developed SGS model automatically satisfies the reflectional invariance and wall boundary conditions without an extra training process so that DRL can quickly find the optimal policy. Furthermore, direct accumulation of reward, spatially and temporally correlated exploration, and the pre-training process are applied for the efficient and effective learning.  In various environments, our DRL could discover SGS models that produce the viscous and Reynolds stress statistics perfectly consistent with the filtered DNS. By comparing various statistics obtained by the trained models and conventional SGS models, we present a possible interpretation of better performance of the DRL model. 
\end{abstract}

\maketitle

%

	\section{Introduction}\label{sec1}
Turbulent flows are observed in many scientific, engineering, and medical problems, such as climate and weather forecasting, design of aircraft, turbines, and pumps, and identification of vascular diseases. For a deep understanding of physics and advanced design in such problems, precise numerical simulation is very important, but the cost is prohibitive because of the nonlinear and multi-scale nature of turbulence. Despite the rapid growth of computational hardware and numerical algorithms, applying direct numerical simulation (DNS) \citep{Moin1998}, which resolves the smallest-scale turbulence, to real-world problems at high Reynolds numbers is expected to be impossible in the near future. For more than half a century, the development of turbulence models to improve the inevitable trade-off between cost and accuracy of simulation has been actively carried out and is still one of the most challenging issues in the turbulence research field.

Most turbulence modeling is categorized into Reynolds-averaged Navier--Stokes (RANS) and large-eddy simulations (LES) (see reviews \citep{Lesieur1996,Meneveau2000,Piomelli2002,Pope2004,Spalart2009,Bose2018,Moser2021}). Unlike RANS, LES, which is the focus of the present work, is time-dependent and scale resolving, making modeling more difficult. In LES, the contribution by sub-grid scale (SGS) motions, which are unknown information, is modeled by resolved-scale flow variables, which can be handled. Many types of algebraic SGS models based on statistical analysis and physical observations have been developed and applied to many practical problems. The most representative model is the Smagorinsky model \citep{Smagorinsky1963}, in which, under the Boussinesq hypothesis, the anisotropic part of the SGS stress tensor is modeled by multiplying the resolved strain-rate tensor $\bar{S}_{ij}$ and scalar eddy-viscosity ${\nu_t}$. In an effort to extend LES to various turbulent flows or complex geometries, some modifications of the Smagorinsky model were proposed. For example, \citet{Germano1991} and \citet{Lilly1992} suggested a dynamic Smagorinsky model (DSM) with an adaptive procedure based on the scale-invariance assumption for determining the Smagorinsky constant ($C_s$) in $\nu_t$. \citet{Vreman2004} developed an efficient model using only first-order derivatives, and it can be applied to wall-bounded flows without any explicit clipping, filtering, averaging, or wall-damping function. As an extension of this research, dynamic models \citep{Park2006,You2007} were also proposed. There are other types of models, such as the scale-similarity model \citep{Bardina1980,Liu1994}, gradient model \citep{Leonard1974,Clark1979}, and mixed model \citep{Zang1993,Salvetti1995,Horiuti1997} (see details in the above review papers). Although SGS models continue to evolve and are subsequently applied to practical problems, there is a strong need for significant improvement in terms of accuracy and cost. As evidence, DSM and the scale-similarity model have provided inaccurate results in coarse grid LES of canonical flows, which implies that these models do not guarantee successful performance in a new environment that has not been verified. In addition, the existing SGS models are not satisfactory with respect to the accuracy of statistics (especially in turbulence intensity), the expression of backscatter (negative eddy-viscosity), the need for ad-hoc methods to prevent numerical instability, and applicability to complex flows relevant to the test-filter operation.

Recently, machine-learning-based models trained using high-fidelity data, mostly DNS, have emerged to address the above issues (see reviews \citep{Kutz2017,Brenner2019,Duraisamy2019,Brunton2020,Duraisamy2021}). SGS models have been developed in several canonical flows, including two-dimensional (2D) homogeneous isotropic turbulence (HIT) \citep{Maulik2017,Maulik2019,Guan2021,Kochkov2021}, three-dimensional (3D) forced HIT \citep{Vollant2017,Zhou2019,Xie2020,Portwood2021,Frezat2021,Prakash2021}, 3D decaying HIT \citep{Wang2018,Beck2019,Prakash2021}, 3D compressible HIT \citep{Xie2019a,Xie2019b}, 3D turbulent channel flow \citep{Wollblad2008,Gamahara2017,Park2021}, and 3D circular cylinder flow \citep{Font2021}. Most of the research is based on the classical supervised (offline) learning framework, which is usually composed of three steps: (i) data collection and training, (ii) \textit{a priori} test, and (iii) \textit{a posteriori} test. In the first step, the model, mostly a deep neural network (DNN), is trained to produce accurate SGS stresses obtained from DNS as the target, and the input of the model is pre-determined among the resolved flow variables such as velocity, velocity gradient, strain rate, rotation rate, second derivative of velocity, and wall-normal height. In the second step, the trained models are quantitatively assessed using data that are similar or slightly different from training data, such as flow fields at different times or at higher Reynolds numbers. In the final step, the actual LES is carried out with the trained model, and its performance is quantified and evaluated. If the \textit{a posteriori} test result is not satisfactory, the cumbersome process must be repeated by changing the input information or network architecture. This drawback comes from the significant gap between \textit{a priori} and \textit{a posteriori} tests. Although the reason for the mismatch is unclear, it is most likely because the model was trained only at the equilibrium state, and the effect (reaction) of the modeled SGS stress in the actual LES was not taken into account in the training process. Another major drawback of this framework is the requirement for high-fidelity data. In real-world problems that require LES, DNS is usually not feasible, and only some statistical data at relatively low Reynolds numbers can be collected. For practical application of the machine-learning-based SGS model, an alternative algorithm is necessary. We expect that deep reinforcement learning (DRL), an online learning algorithm, would help overcome the disadvantages of classical supervised learning.

DRL, in which the training process and simulations (or experiments) are carried out jointly, has been drawing attention as a promising tool for discovering new schemes for control in fluid mechanics. Some application examples include efficient swimming \citep{Colabrese2017,Verma2018}, drag reduction of flow around a cylinder \citep{Rabault2019,Rabault2019a,Tang2020,Fan2020,Paris2021}, heat transfer control \citep{Beintema2020,Hachem2021}, control of chaotic systems \citep{Bucci2019,Zeng2021}, and shape optimization \citep{Viquerat2021}, with recent reviews \citep{Rabault2020,Ren2020,Garnier2021}. Most of these studies were carried out in the laminar flow regime, and the DRL of turbulent flow has rarely been attempted and thus remains a challenging problem. In particular, DRL for turbulence modeling is more challenging because it requires an optimization of the SGS stress at every grid point; that is, the action space is very high-dimensional. In the present work, we employed DRL for SGS modeling for a large-eddy simulation of turbulent flow. Highly relevant to our study, \citet{Novati2021} first proposed the DRL-LES framework for developing SGS models in 3D forced HIT and reported that the model trained with statistics of energy spectrum as a target has better statistical accuracy and numerical stability than the conventional model, DSM. They demonstrated the generalization ability of the model in terms of the Reynolds number effect. However, the learning was not successful in the case of fully dispersed agents at all grids, and only SGS models based on the linear eddy-viscosity assumption using the Smagorinsky-type form were considered. DRL has not yet been applied to LES for other types of flows. For a similar purpose, although not DRL, the adjoint-based supervised learning framework was proposed by \citet{Sirignano2020} as an \textit{a posteriori} model training algorithm that uses the Navier--Stokes equations to reflect the temporal reaction of SGS stress. \citet{MacArt2021} generalized this framework to use only statistical information as the training target. However, it required the DNS flow fields, which are generally not available, as initial conditions for solving adjoint equations, and only the case of using statistics of the full-time horizon for the training target in developing flows was considered. Furthermore, training through automatic differentiation of Navier--Stokes equations can be unstable when unrolling over multiple time steps \citep{Kochkov2021}. 

\begin{figure*}
	\centerline{\includegraphics[width=1.6\columnwidth]{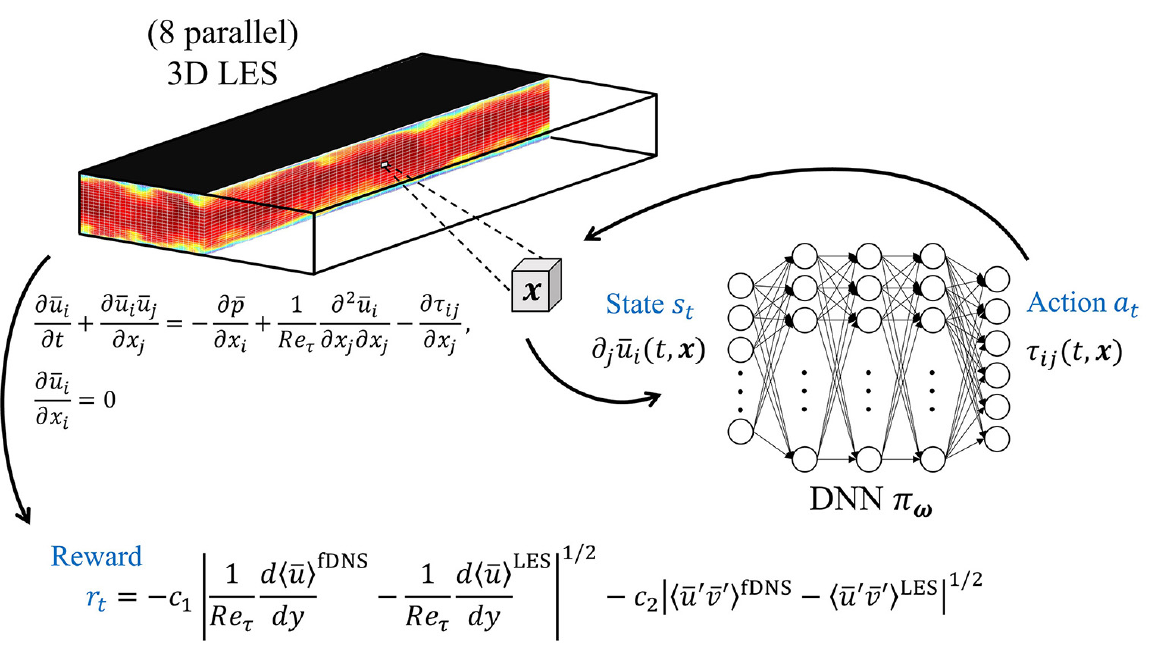}}
	\caption{Illustration of present work. A DRL-LES framework for developing a subgrid-scale (SGS) model in wall-bounded turbulence was proposed. Running of three-dimensional (3D) LES and learning of deep neural network (DNN) that produces the SGS stress or the eddy viscosity from the resolved velocity gradient are carried out simultaneously. The target statistics for training are the mean viscous stress and mean Reynolds shear stress, which are used for reward calculation.}
	\label{fig01}
\end{figure*}

In this study, we extend the DRL-LES framework to the development of an SGS model for wall-bounded turbulence (Fig. \ref{fig01}). To overcome these limitations, we propose a physics-constrained DRL algorithm that perfectly guarantees the reflectional equivariance and boundary conditions of the SGS model. This makes the possible solution space narrow and thus significantly reduces the cost of DRL. For more efficient and effective learning of turbulence, we suggest several techniques including reward localization, N-step actor-critic, parameter space exploration with the Ornstein--Uhlenbeck (OU) process, and pre-training of the policy network. In addition, a simplified algorithm with similar performance is presented for easy implementation through hyper-parameter reduction. We successfully found optimal SGS stresses with a very high dimension $(> 10^5)$ and inhomogeneous distribution by the wall effect. Finally, we analyze DRL models through various statistics compared with conventional SGS models and discuss the difference between DRL models with and without the eddy-viscosity assumption. Section \ref{sec2}, Section \ref{sec3}, and Section \ref{sec4} consist of the methodology, results of DRL, and conclusions, respectively.

\section{Methodology}\label{sec2}
This section includes the governing equations and numerical methods of the LES and DRL algorithms and important techniques for the DRL of turbulent flows. 

\subsection{Large-eddy simulation of wall-bounded turbulence}
The governing equations of the DNS of turbulent channel flows, which are necessary for the extraction of filtered data for the target statistics in later learning, are the continuity and incompressible Navier--Stokes equations as follows:
\begin{equation}
	{\partial u_i \over \partial x_i} = 0,
\end{equation}
\begin{equation}
	{\partial {u_i} \over \partial t} + {\partial (u_i u_j) \over \partial x_j} = -{\partial p \over \partial x_i} + {1 \over Re_\tau}{\partial^2{u_i} \over \partial x_j \partial x_j}.
\end{equation}
The equations are non-dimensionalized by the length scale of the channel half width ($\delta$) and friction velocity $u_\tau$. $x_1$ ($x$), $x_2$ ($y$), and $x_3$ ($z$) denote the streamwise, wall-normal, and spanwise directions, respectively, and the corresponding velocity components are $u_i (=u,v,w)$. The major parameter is the friction Reynolds number $Re_\tau={u_\tau \delta \over \nu}$, where $\nu$ is the kinematic viscosity. Three DNSs at different $Re_\tau$ were carried out to collect the target-filtered statistics and the evaluation of actual LESs, as given in Table \ref{t.DNS}. The DNSs of $Re_\tau=180,360$ are used for both training and testing, whereas that of $Re_\tau=720$ is used only for comparison with the test results.

\begin{table}
	\begin{center}
		\begin{tabular}{cccccc}
			& $Re_\tau$ & ($L_x$, $L_y$, $L_z$)  & ($N_x$, $N_y$, $N_z$) & ($\Delta x^+$, $\Delta z^+$) \\ \hline
			& 180 & ($4\pi, 2, 2\pi$)  &(128, 129, 128) & (17.7, 8.84) \\[1mm]
			& 360 & ($2\pi, 2, \pi$) & (128, 129, 128) & (17.7, 8.84) \\[1mm]
			& 720 & ($\pi, 2, 0.5\pi$) & (128, 193, 128) & (17.7, 8.84) \\[1mm]
			
		\end{tabular}
		\caption{DNS setups for collecting target statistics of DRL. $L_x$, $L_y$, and $L_z$ are the domain lengths normalized by the channel half width in the streamwise, wall-normal, and spanwise directions, respectively. $N_x$ and $N_z$ are the number of Fourier modes in the homogeneous directions, and $N_y$ is the number of wall-normal grids. $\Delta x^+$ and $\Delta z^+$ are the grid sizes of the wall units. In all cases, the minimum size of the wall-normal grid in the wall units is smaller than 0.25.}{\label{t.DNS}}
	\end{center}
\end{table}

The governing equations of the LES are obtained by a spatial filter operation ($\bar{\cdot}$) on the above equations,
\begin{equation}
	{\partial \bar{u}_i \over \partial x_i} = 0,
\end{equation}
\begin{equation}
	{\partial {\bar{u}_i} \over \partial t} + {\partial  (\bar{u}_i \bar{u}_j) \over \partial x_j} = -{\partial \bar{p} \over \partial x_i} + {1 \over Re_\tau}{\partial^2{\bar{u}_i} \over \partial x_j \partial x_j} - {\partial{\tau_{ij}}\over \partial x_j},
\end{equation}
where $\tau_{ij}$ is the SGS stress that should be modeled. For the spatial discretization, a pseudo-spectral method is used in the homogeneous (streamwise and spanwise) directions, and a second-order central difference scheme in nonuniform grids is used in the wall-normal direction. For time integration, the second-order Adams--Bashforth method is used for the nonlinear and SGS terms, and the Crank--Nicolson method is used for the viscous term. 

In this study, we model $\tau_{ij}$ using a deep neural network (DNN, $\pi_\omega$) with trainable parameters $\omega$ that can represent a highly nonlinear relation using the local resolved velocity gradient $\bar{\alpha}_{ij}(=\partial \bar{u}_j / \partial x_i)$ as input, that is, $\tau_{ij} \approx \pi_\omega(\bar{\alpha}_{ij})$. In supervised learning, $\pi_\omega$ is trained using the filtered DNS (fDNS) flow field ($\tau_{ij}^{\mathrm{fDNS}} = \overline{u_iu_j}-\bar{u}_i\bar{u}_j$) as the target. However, this approach has limitations, such as the inaccessibility of DNS data and inaccuracy in the \textit{a posteriori} test, as mentioned in Section \ref{sec1}. In the present work, we trained $\pi_\omega$ via DRL, a class of online learning algorithms, without using the fDNS flow field. The objective function of DRL for SGS modeling to be minimized is the distance between the LES statistics ($S^{\mathrm{LES}(\pi_\omega)}$) and high-fidelity statistics (e.g., $S^{\mathrm{fDNS}}$), and it can be defined as follows:
\begin{equation}\label{eq5}
	\mathrm{argmin}_{\omega} ||S^{\mathrm{target}}-S^{\mathrm{LES}(\pi_\omega)}||.
\end{equation}

Although the ultimate goal of DRL is to find $\pi_\omega$ to achieve the above requirement, its results can be significantly different depending on the choice of state and action, the definition of reward, and some important algorithm techniques. In the sense that the performance of the \textit{a posteriori} test is guaranteed if the training is successful, the construction of a robust learning algorithm is essential. The detailed DRL methods for turbulence modeling are presented in the following subsections.

To assess our DRL model, we used conventional algebraic models widely employed in wall-bounded turbulence, the static Vreman model (SVM) \citep{Vreman2004}, and DSM \citep{Germano1991,Lilly1992}. For both models, the anisotropic part of the SGS stress is represented by the scalar eddy-viscosity and resolved strain-rate tensor, that is, $\tau_{ij}-\tau_{kk}\delta_{ij}/3 = - 2\nu_t \bar{S}_{ij}$ with $\bar{S}_{ij}=(\partial \bar{u}_i / \partial x_j + \partial \bar{u}_j / \partial x_i)/2$. For SVM, $\nu_t = 2.5C_s^2 \sqrt{{B_\beta}/ (\bar{\alpha}_{ij}\bar{\alpha}_{ij})}$ where $\beta_{ij} = \bar{\Delta}_m^2\bar{\alpha}_{mi}\bar{\alpha}_{mj}$ with the m-directional grid size $\bar{\Delta}_m$, $B_\beta=\beta_{11}\beta_{22}-\beta_{12}^2+\beta_{11}\beta_{33}-\beta_{13}^2+\beta_{22}\beta_{33}-\beta_{23}^2$. This model requires only the first derivative of velocity without an ad hoc method or average operation; thus, the computational cost is low. For DSM, $\nu_t = C^2|\bar{S}|$ with dynamically determined $C^2 =\langle L_{ij}M_{ij}\rangle_h / \langle 2M_{ij}M_{ij}\rangle_h$. Here, $|\bar{S}|=\sqrt{2\bar{S}_{ij}\bar{S}_{ij}}, L_{ij}=\widetilde{\bar{u}_i\bar{u}_j}-\tilde{\bar{u}}_i\tilde{\bar{u}}_j, M_{ij}= \widetilde{|\bar{S}|\bar{S}_{ij}}-(\tilde{\Delta}/\bar{\Delta})^2 |\tilde{\bar{S}}|\tilde{\bar{S}}_{ij}$, $\langle \rangle_h$ denotes an average operation in the horizontal directions, and $\bar{\Delta}$ and $\tilde{\Delta}(=2\bar{\Delta})$ are the size of the grid ($\bar{\cdot}$) and test filters ($\tilde{\cdot}$), respectively. This model achieves numerical stability through an average operation, eliminating negative $\nu_t$. A similarity model is not considered for comparison because the DSM usually outperforms it.

\subsection{Reinforcement learning algorithm: deep deterministic policy gradient}
Reinforcement learning (RL) is a learning algorithm that finds the optimal action $a_t$ of an agent to receive the maximum reward in a given state $s_t$ of the environment. Here, the target to maximize is not the instantaneous reward $r_t$ coming out after the action but the long-term reward ($R_t = \sum_{i=t}^{\infty} {r_i}$) or discounted long-term reward ($R_t = \sum_{i=t}^{\infty} {\gamma^{i-t} r_i}$ with $0<\gamma<1$) that is accumulated over a time horizon. For example, in our problem, the environment is LES, and the state and action are the resolved flow variables and modeled ones of LES, respectively, and the instantaneous reward can be defined as the statistical accuracy of the LES after a short time period. In the training process, the RL algorithm optimizes the SGS model in the direction of maximizing the long-term statistical accuracy of the LES (see Fig. \ref{fig01}).

\begin{figure*}
	\centerline{\includegraphics[width=1.8\columnwidth]{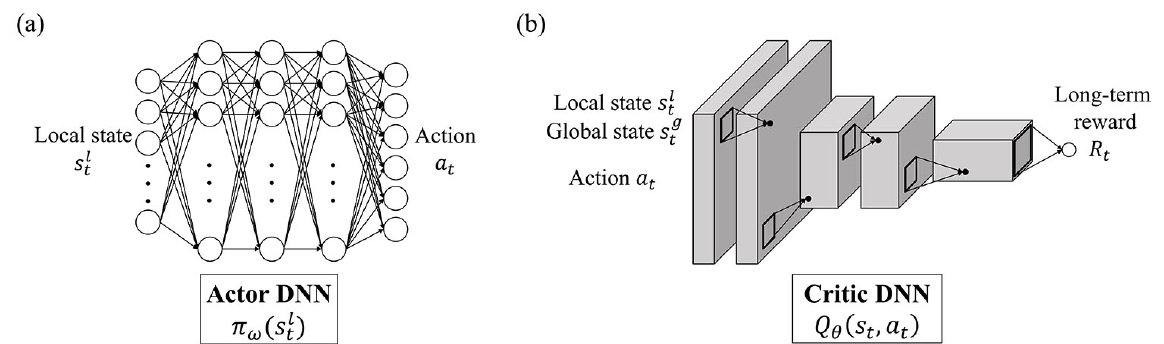}}
	\caption{Actor-critic algorithm composed of two DNNs. The actor DNN is a fully connected network that produces local SGS stresses based on locally resolved information, whereas the critic DNN is a convolutional neural network that predicts long-term reward based on the local and global states and corresponding action.}
	\label{fig02}
\end{figure*}

For current turbulence modeling that requires a nonlinear and continuous model, the deep deterministic policy gradient algorithm (DDPG) is employed as a DRL algorithm \citep{Lillicrap2015}. The DDPG algorithm is a class of the actor-critic algorithm that is composed of critic DNN and actor DNN that can express the action in continuous space. As a surrogate model for the environment, a critic DNN (action-value function $Q_\theta$ with parameter $\theta$) approximates the long-term reward based on the given state and action:	
\begin{equation}
	\mathbb{E}_{\pi_\omega} \left[\sum_{i=t}^{\infty} {\gamma^{i-t} r_i}\right] \approx Q_\theta (s_t,a_t) .
\end{equation}	
Here, the training of $Q_\theta$ with deterministic policy mostly depends on a recursive relation of the Bellman equation:	
\begin{equation}\label{eq7}
	Q_\theta (s_t,a_t) = r_t + \gamma Q_\theta (s_{t+1},\pi_\omega(s_{t+1})).
\end{equation}	
From this, we can obtain the objective function of parameter $\theta$ in the critic DNN:	
\begin{equation}\label{eq8}
	\mathrm{argmin}_{\theta} \mathbb{E}_{s_t, a_t, r_t, s_{t+1}} \left[(Q_\theta (s_t,a_t)-y_t)^2\right] ,
\end{equation}
where 
\begin{equation}\label{eq9}
	y_t = r_t + \gamma Q_{\theta^*} (s_{t+1},\pi_{\omega^*}(s_{t+1})) .
\end{equation}	
Here, for the calculation of $y_t$, target networks $Q_{\theta^*}$ and $\pi_{\omega^*}$ are usually used to avoid the oscillation of training \citep{Mnih2013,Mnih2015}, and a soft update method of target ones is a widely used approach where $\theta^* \leftarrow (1-\eta)\theta^* + \eta \theta$ and $\omega^* \leftarrow (1-\eta)\omega^* + \eta \omega$ with $\eta \ll 1$. For training, datasets $\left\lbrace s_t, a_t, r_t, s_{t+1}\right\rbrace$ from the environment are required and used to be memorized in the replay buffer ($B^{DRL}$) for sampling efficiency. 

Finally, using a critic DNN that predicts the long-term reward, we can train the actor DNN $\pi_{\omega}$ in the direction of increasing critic value. The objective function to maximize is as follows:	
\begin{equation}\label{eq10}
	\mathrm{argmax}_{\omega} \mathbb{E}_{s_t} \left[Q_\theta (s_t,\pi_{\omega}(s_t))\right] .
\end{equation}	
As shown in Fig. \ref{fig02}, in our study, the actor DNN is composed of fully connected layers, whereas the critic DNN is a convolutional neural network. In DRL, the data collection of $\left\lbrace s_t, a_t, r_t, s_{t+1}\right\rbrace$ from running the environment with policy $\pi_{\omega}$ and the learning of $Q_\theta$ and $\pi_{\omega}$ based on Eqs. (\ref{eq8}) and (\ref{eq10}) is repeated. As a result, it is expected that the ultimate goal of the DRL finding optimal $\pi_{\omega}$ is achieved. Unfortunately, DRL can be highly unstable depending on the given environment, and the critical reason is presumed to be due to the error of the value function approximation through DNN, as addressed by \citet{Fujimoto2018}. They proposed the twin delayed DDPG (TD3) with some techniques, including clipped double Q learning, delayed policy update, and target policy smoothing, which are state-of-the-art algorithms. We employ this algorithm as a baseline, and for the successful application of DRL to turbulent flows, we suggest some modifications as given in the following subsections.

\subsection{State and action information and definition of reward function} \label{sec2.3}
The results and cost of DRL are highly dependent on the setting of the state, action, and reward. Related issues were considered in previous studies on the RL of fluid mechanics. \citet{Rabault2019} reported that drag reduction performance in a cylinder flow depends on the amount of state information. To overcome the difficulty of handling 3D data as the state in LES, \citet{Novati2021} presented a multi-agent RL framework that performs localized actuation based on local and global flow information. \citet{Fan2020} reported that the stability of learning varies depending on the selection of auxiliary reward or the use of a filtering process. Therefore, careful design is required in the construction of a successful RL framework.

In this study, the 3D whole flow data of LES can be used as state information. However, storing such data in the relay buffer and constructing the critic DNN of a 3D convolutional neural network (CNN) is expensive and becomes impossible as the grid resolution increases. Therefore, we use both local information ($s_t^l(x,y,z)$) including local velocity gradient $\bar{\alpha}^+_{ij}(x,y,z)$ and local grid size $\bar{\Delta}^+ (=(\bar{\Delta}^+_x\bar{\Delta}^+_y\bar{\Delta}^+_z)^{1/3})$, which are normalized in wall units, and global (statistical) information ($s_t^g(y)$) including mean velocities ($\left< \bar{u} \right>$, $\left< \bar{w} \right>$), mean shear stresses ($\left< \partial \bar{u}/\partial y \right>, \left< \partial \bar{w}/\partial y \right>$), root-mean-square (rms) of the velocities ($\bar{u}_{i,rms}$), mean Reynolds shear stress ($\left< \bar{u}'\bar{v}' \right>$), and wall-normal locations ($y^+$), averaged in the homogeneous direction (Fig. \ref{fig02}). For a more accurate long-term reward estimation, local information from several grids on the horizontal plane rather than one grid is used. We confirmed that the variance is sufficiently low, and the learning performance is good enough when using more than $8\times1\times8$ grids (in $x \times y \times z$), thus we used $8\times1\times8$ patch information to learn the critic DNN.

As the input of the actor DNN, the local information at the point of interest $s_t^l$, which is part of the state information, is used to predict the local SGS stress at that point. If statistical information is additionally used in the actor DNN, there is a possibility of overfitting in the learning environment, resulting in the disadvantage that an average operation is required to calculate the SGS stress after learning. To save computational cost in LES modeling and scalability with complex geometry, it is advantageous to use only local information for the actor DNN. Similar to the SVM, only the velocity gradient $\alpha_{ij}$ and grid size $\bar{\Delta}$ are used as the input, and the second derivative information is not considered for cost savings. Two cases are considered for the output of the actor DNN: (i) a scalar $\nu_t$ with a linear eddy-viscosity assumption, in which $\tau_{ij}-\tau_{kk}\delta_{ij}/3 = - 2\nu_t \bar{S}_{ij}$ and (ii) a tensor $\tau_{ij}$, which is composed of six components considering the symmetry of the tensor. However, the $\nu_t$ model yields degraded performance owing to the constraint that the SGS stress tensor is proportional to the strain-rate tensor, as shown in Section \ref{sec3-3}. We mainly focused on the $\tau_{ij}$ case, which has higher degrees of freedom. In addition, a DNN whereby the filter (grid) size is not considered is likely to produce non-physical SGS stress. For example, even when the grid resolution is very fine, the DNN can produce a strong SGS stress, which is supposed to vanish in DNS resolution. Therefore, we added an operation that multiplies the output of the last layer in the DNN by the square of the filter size, similar to conventional SGS models. We found that the DNN with consideration of the filter size has better generalization performance with respect to grid resolution than the case without consideration.

The reward function for LES modeling can be defined as statistical accuracy. While there are various candidates for the statistics, such as the mean velocity, rms of the velocity fluctuations, and energy spectrum, we consider the total shear stress equation:	
\begin{equation}\label{eq11}
	\left<{1\over Re_\tau} {d\bar{u}\over dy}\right> - \left< \bar{u}'\bar{v}'\right> - \left<\tau_{xy}\right> = 1 - y,
\end{equation}
where ${1\over Re_\tau} {d\bar{u}\over dy}$, $\bar{u}'\bar{v}'$, and $\tau_{xy}$ are the viscous stress, Reynolds stress, and SGS shear stress, respectively. The use of above equation has several advantages. First, the above statistics are localized at a specific wall-normal height, and the state and action at the height are predominantly related to them. Second, in a fully developed turbulent channel flow, if only two quantities among $\left<{1\over Re_\tau} {d\bar{u}\over dy}\right>$, $\left< \bar{u}'\bar{v}'\right>$ and $\left<\tau_{xy}\right>$ are matched with targets, the remaining one is automatically determined from the global balance. Third, each stress has a similar order of magnitude, and thus, it is intuitive to consider the relative weights. We tested two cases: (i) accuracy of $\left<{1\over Re_\tau} {d\bar{u}\over dy}\right>$ and $\left< \bar{u}'\bar{v }'\right>$, and (ii) accuracy of $\left<{1\over Re_\tau} {d\bar{u}\over dy}\right>$ and $\left<\tau_{xy}\right> $. We found that DRL for both cases was successful, although the latter case showed slightly more stable training. However, one might think that $\left<\tau_{xy}\right> $ is not available in real-world applications. Therefore, we considered only the former case in this study. 

More importantly, reward localization in the wall-normal direction was applied. The accuracy of stress statistics at a specific height is poorly correlated with remote SGS stresses and is predominantly affected by nearby ones. The reward defined with global accuracy across the whole domain could make it difficult to judge which action is good or bad, whereas the reward defined with local accuracy can provide a more direct guide to local actuation. The localized instantaneous reward function is defined as 	
\begin{equation}\label{eq12}
	r_t(y) = \sum_{i=1}^2 -c_i \left|S_i^\textrm{fDNS} (y) - S_i^\textrm{LES} (t+\Delta t^\textrm{DRL},y) \right|^{1/2},
\end{equation}
where 
\begin{equation}
	S_1 = \left<{1\over Re_\tau} {d\bar{u}\over dy}\right>, ~~~S_2 = \left< \bar{u}'\bar{v}'\right>. 
\end{equation}
Here, the coefficients $c_i$ were chosen to be of order 1 for the purpose that both rewards are properly maximized. When $c_1=c_2 = 1$, however, the mean velocity relevant with $S_1$ fluctuated frequently and learning was quite unstable. It was conjectured that the temporal reaction scale of $S_1$ and $S_2$ are different. Therefore, we adjusted $c_1=1$ and $c_2 = 0.5$, which yielded the best performance in the learning process. The target statistics $S_i^\textrm{fDNS}$ are time-averaged statistics of the filtered DNS data. Although only the fDNS statistics are considered as targets in this study, it is possible to select other statistics depending on the accessibility of information such as statistics of the DNS, high-fidelity LES, and even experiments. $\Delta t^\textrm{DRL}$ of the LES statistics $S_i^\textrm{LES} (t+\Delta t^\textrm{DRL})$ is the data collection and learning period of the current DRL and was chosen to be the minimum number of time steps that warrant a statistical variety of learning data. The optimum value is 30 simulation time steps, which corresponds to 5.4 wall time units.  Furthermore, in the definition of reward, the square root is applied to reduce the difference of reward scale in the different wall-normal locations, and this choice yields better training performance than using the absolute value.

\subsection{Effective and efficient DRL techniques}\label{sec2-4}

In this section, we propose several techniques for the effective and efficient learning of turbulent flows that relieve the oscillation of training and accelerate the training speed. The key techniques include the N-step reward, spatiotemporally correlated and state-dependent exploration, and the pretraining of the policy network, as follows.

First, the N-step technique combined with a time-average operation is used to reduce the high bias error in the value function approximation \citep{Barth-Maron2018}. This idea comes from our estimate for the reason of unstable learning that the short-term reward after $\Delta t^\textrm{DRL}$ is insufficient to reflect the long-term evolution of turbulent flows as influenced by the given action, and its wild fluctuations make it difficult to judge whether the proposed action is good or bad. Theoretically, the DRL is not affected by fluctuations because the estimated long-term reward contains the concept of time average. However, in turbulent flows, learning based on the Bellman equation with short-term rewards can be very unstable. Therefore, we apply the technique to consider a future reward and a weighted-time-averaged operation. The objective function of the critic DNN is changed to:
\begin{equation}\label{eq13}
	\mathrm{argmin}_{\theta} \mathbb{E}_{s_t, a_t, r_t, s_{t+N^s}} \left[(Q_\theta (s_t,a_t)-y_t)^2\right] ,
\end{equation}
where 
\begin{equation}
	y_t = r_t + \gamma^{N^s} Q_{\theta^*}(s_{t+N^s}, \pi_{\omega^*}(s_{t+N^s})),
\end{equation}
\begin{equation}\label{eq15}
	r_t = \sum_i -c_i \left|S_i^\textrm{fDNS}- S_i^\textrm{LES*} (t) \right|^{1/2},
\end{equation}
\begin{equation}
	S_i^\textrm{LES*} (t) = {\sum_{j=1}^{N^s} \gamma^{j-1} S_i^\textrm{LES} (t+j\Delta t^\textrm{DRL}) \over \sum_{j=1}^{N^s} \gamma^{j-1}} .
\end{equation}
When $N^s = 1$, Eq. (\ref{eq15}) reduces to the short-time reward Eq. (\ref{eq12}). As $N^s$ increases, more future simulation results are directly reflected in the training. However, an excessively large $N^s$ would cause a large variance, and simulation results irrelevant to the action are excessively reflected. The optimal settings of $N^s$ and $\gamma$ are related to the reaction time scale of the turbulent flow. The effect of the N-step reward is discussed in the following section.

Second, a new exploration method that considers spatio-temporal correlation is proposed. The DRL results are highly dependent on the exploration methods. For the DRL of turbulent flow, Gaussian noise is not effective because spatio-temporally uncorrelated information (very small time-scale) disappears very quickly in the time-integration of the Navier--Stokes equation. In addition, zero-centered Gaussian noise, which is usually used, hardly affects the average value of the action. In channel flow, the scale of optimal action is highly dependent on the wall-normal height, but simple noise cannot reflect this.

Thus, a new spatiotemporally correlated exploration method was developed based on parameter space exploration \citep{Plappert2017} and the OU process \citep{Lillicrap2015, Bucci2019}. The parameter space noise enables more diverse exploration than Gaussian noise and can express a spatially correlated change of action owing to the state dependency. By combining it with the OU process for temporal correlation, we can achieve the reflection of the momentum effect and physical compatibility in the incompressible flow solver. This method can be described as follows:
\begin{equation}
	\tau_{ij} = \pi_{\omega^\epsilon}(s^l) ,
\end{equation}
where	
\begin{equation}\label{eq18}
	\omega_h^\epsilon = \omega_h + \epsilon_h,
\end{equation}
\begin{equation}\label{eq19}
	\epsilon_h \leftarrow (1-\zeta)\epsilon_h + \sigma_{\omega_h} \mathcal{N}(0,\sigma_\epsilon) ,
\end{equation}
where $\omega_h$ is the trainable parameter in the $h$-th layer of the actor DNN, and $\epsilon_h$ is the added noise. $\sigma_{\omega_h}$ is the standard deviation of $\omega_h$ to provide noise with a proper scale in each layer so that trainable parameters are changed evenly across all layers. A scalar constant $\zeta = 0.001$ and $\mathcal{N}(0,\sigma_\epsilon)$ is Gaussian noise with zero mean and a standard deviation of $\sigma_\epsilon$. Here, a scalar variable $\sigma_\epsilon$ is adjusted to maintain a constant distance between $\pi_{\omega}$ and $\pi_{\omega^\epsilon}$. At every DRL step, when the distance calculated through the sampled batch data is greater/smaller than a target distance, $\sigma_\epsilon \leftarrow \sigma_\epsilon/1.01 $ or $\sigma_\epsilon \leftarrow 1.01\sigma_\epsilon$. Furthermore, when the distance was greater than the target, Gaussian noise was not added to prevent the distance from being too large, that is, $\epsilon_h \leftarrow (1-\zeta)\epsilon_h$ instead of Eq. (\ref{eq19}). For the metric, the distance of the correlation coefficient $1-{1 \over K}\sum_k R_k$ is used, where $R_k=cov(\pi_{k,\omega},\pi_{k,\omega_\epsilon})/(\sigma_{\pi_{k,\omega}}\sigma_{\pi_{k,\omega_\epsilon}})$, and $R_k$, $cov$, $k$, and $K$ are the correlation coefficient, covariance, component index, and the number of components, respectively. The normalized root-mean-square value can also be used as an alternative. We found that the proposed noise is effective for speeding up training and obtaining a better solution. The comparison results of the proposed exploration noise and decorrelated Gaussian noise are presented in the following section.

A challenging aspect of DRL for LES in wall-bounded turbulence is that we do not know the proper scale of the SGS model, which is the output of the actor DNN. Commonly used random initialization of actor DNNs would make numerical simulation easily diverge or training may take too long. To overcome this problem, we propose a simple pretraining method using conventional SGS models. Although the accuracy of these models is not satisfactory, it is sufficient for initializing the proper scale of the actor DNN. The objective function for the pretraining is as follows:
\begin{equation}
	\mathrm{argmin}_{\omega} \mathbb{E}_{s^l} [ ||\tau_{ij}^\textrm{SL}-\pi_{\omega}(s^l) ||^2_2] .
\end{equation}
Since an LES with a conventional SGS model is carried out, datasets of input $s^l$ and target $\tau_{ij}^\textrm{SL}$ are collected and then used for the supervised learning. Because only a short simulation is sufficient for the training, this does not significantly affect the overall computational cost of the DRL framework. In contrast, the overall training cost can be significantly reduced by using pretraining. We assessed the differences caused by the different conventional SGS models used for pretraining. For example, when using SVM and DSM, successful optimal models were found by DRL, although the SVM case gave slightly faster convergence than the DSM. On the other hand, when using the scale-similarity model for pretraining, DRL is unstable owing to the excessive numerical instability of LES, suggesting that a clipping operation is necessary. To achieve stable learning without using a clipping operation, it might be necessary to consider the complex design of the reward function and data processing relevant to the blow-up of LES, which we did not consider in this study. Although SVM was used for most pretraining in this work, we expect that other types of eddy-viscosity models might be good alternatives. By applying the above three key techniques using the DRL framework, we successfully developed the LES model for wall-bounded turbulence, and we found that the characteristics of the trained model could be different depending on pretraining.

\subsection{Physical constraints on neural network: reflectional equivariance and wall boundary condition}\label{sec2-5}

\begin{figure*}
	\centerline{\includegraphics[width=1.2\columnwidth]{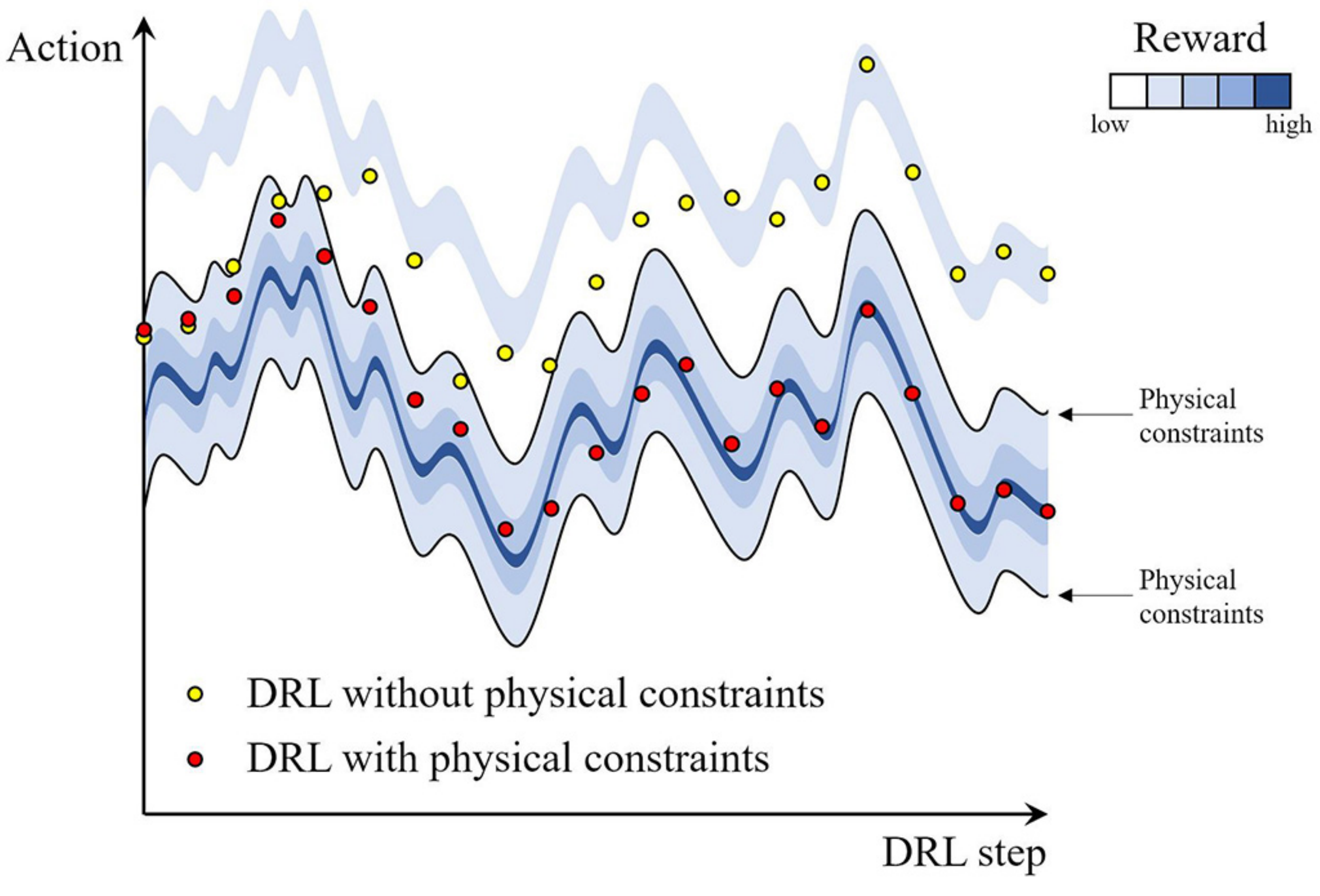}}
	\caption{Illustration of DRL with physical constraints in turbulent flows. A DRL-LES framework for developing a subgrid-scale (SGS) model for wall-bounded turbulence was proposed. DRL without physical constraints violates physical laws in the learning process, while DRL with physical constraints explores optimal action, which gives the highest reward, within the physical laws.}
	\label{fig03}
\end{figure*}

In this section, we propose a method to apply physical constraints to the DRL to dramatically reduce cost. The expected scenario (Fig. \ref{fig03}) is that the constraints prevent unphysical exploration and thus help the DRL reach an optimal solution quickly. It is known that a pure NN trained with limited data cannot accurately reflect the physical laws, including Galilean invariance, rotational invariance, reflectional invariance, unit invariance, and boundary conditions. Through many attempts to satisfy these properties in supervised learning, some successful methods have been developed. For example, a well-known method is the tensor-based NN for Galilean invariance and rotational invariance, which has been successfully applied to RANS modeling \citep{Ling2016}. However, the number of tensors is large and computational complexity exists. Other ways to reflect the physical laws are to perform data augmentation using invariance \citep{Kim2020,Kim2020b,Frezat2021}, adding constraints to the loss function \citep{Raissi2019, Lee2019}, or adding an additional reward term usually used in RL \citep{Rabault2019,Fan2020}. However, these approaches can yield imperfect invariance and require hyperparameter tuning. Even when well-tuned, the physical law can be broken before sufficient training. 

Turbulent channel flow has statistical reflectional symmetry in the wall-normal and spanwise directions. It is important to satisfy the reflectional equivariance in the wall-normal direction, $\pi_\omega(s^l)= \rho^{2-y}(\pi_\omega(\rho^y(s^l)))$, and the spanwise direction, $\pi_\omega(s^l)= \rho^{-z}(\pi_\omega(\rho^z(s^l)))$, where $\rho$ is the reflection (mirroring) operation. If the equivariant conditions are satisfied, the generation of unphysical flow, such as non-negligible mean flow in the spanwise direction, can be prevented.

The reflectional equivariance can be achieved by a simple operation that uses $\pi^r_\omega(s^l) = (\pi_\omega(s^l)+ \rho^{2-y}(\pi_\omega(\rho^y(s^l))))/2$ instead of $\pi_\omega(s^l)$ in the wall-normal direction. It is obvious that $\pi^r_\omega$ automatically satisfies the reflectional property. Applying spanwise reflectional equivariance is straightforward using the same technique. This approach is applied not only to the actor DNN, but also on the critic DNN, which is expected to return the same value from the mirrored data. 

For the zero SGS stress condition ($\pi_\omega(s^l)|_{y=0,2} = 0$) at the wall boundary, a similar approach can be applied. Under the no-slip and divergence-free conditions, it is obvious that seven velocity gradient components other than $\partial_yu$ and $\partial_yw$ are zero at the wall. With the seven components denoted as $s^{l*}$, we can modify the actor DNN as $\pi_\omega^{b} = ||s^{l*}||_2(\pi_\omega(s^l))$. $\pi_\omega^{b}$ automatically satisfies the zero SGS stress when $s^{l*}$ is zero. We emphasize that an {\it ad hoc} function, such as the van Driest damping, is not used. The application of the same method to other input information (such as the second derivative of velocity) is straightforward. Finally, the actor DNN $\pi_\omega^{r,b}$ satisfying the physical constraints of the reflectional equivariance and the boundary condition can be constructed as 
\begin{multline}
	\pi_\omega^{r,b} = ||s^{l*}||_2(\pi_\omega(s^l) + \rho^{2-y}(\pi_\omega(\rho^y(s^l))) \\
	+ \rho^{-z}(\pi_\omega(\rho^z(s^l))) + \rho^{2-y,-z}(\pi_\omega(\rho^{y,z}(s^l)))/4 .
\end{multline}
The superscript $^{r,b}$ is omitted hereinafter.

By reflecting the physical properties of the DNN, unnecessary learning processes (e.g., occurrence of the spanwise mean velocity, production of high SGS stresses very near the wall, unphysical exploration, etc.) could be prevented, leading to a cost reduction in learning and improvement in performance. Some modifications to constrain physical law are possible, and a similar study was also conducted through a symmetry reduction process of DRL to control chaotic dynamics \citep{Zeng2021}. We also tested tensor basis NN \citep{Ling2016} to imbed rotational invariance, but unfortunately the extra constraints did not conclusively help improve DRL performance, presumably due to strong anisotropy of the near-wall turbulence. Although the tensor invariance property might be an essential element for the development of a general model, our model only covers the reflectional invariance, which is related to symmetrical statistics of channel flow. Finally, we emphasize that the remaining amount to train the SGS model through DRL is still considerable because the LES prediction accuracy of conventional SGS models, which already reflects those physical properties, is low. 

\subsection{Simplified algorithm and implementation}

In this study, we present a simplified DRL algorithm for easy implementation and reduction of the number of hyperparameters. As presented in the next section, learning based on the recursive relation of the Bellman Eq. (\ref{eq13}) has no significant effect on turbulent flow (the results are given in Section \ref{sec3}), and it is confirmed that only using the truncated time-averaged reward, as given in Eq. (\ref{eq22}), is sufficient for stable and steady training.
\begin{equation}\label{eq22}
	\mathrm{argmin}_{\theta} \mathbb{E}_{s_t, a_t, r_t} \left[(Q_\theta (s_t,a_t)-r_t)^2\right] ,
\end{equation}
where 
\begin{equation}\label{eq23}
	r_t = \sum_i -c_i \left|S_i^\textrm{fDNS}- S_i^\textrm{LES*} (t) \right|^{1/2},
\end{equation}
\begin{equation}\label{eq24}
	S_i^\textrm{LES*} (t) = {\sum_{j=1}^{N^s} \gamma^{j-1} S_i^\textrm{LES} (t+\Delta t^\textrm{DRL}j) \over \sum_{j=1}^{N^s} \gamma^{j-1}} .
\end{equation}
Therefore, by omitting related parts such as storage of the next state, construction of the target DNN, and calculation of the target value, the algorithm can be simplified, and the pseudocode is summarized in Algorithm \ref{Alg1}.

\begin{algorithm}[H]
	\caption{Simplified deep reinforcement learning for turbulent flows}\label{Alg1}
	\begin{algorithmic}[1]
		\State Initialize actor network $\pi_\omega$ and critic network $Q_\theta$ with random parameters $\omega, \theta$
		\State Initialize replay buffers for supervised learning ($B^\textrm{SL}$) and for deep reinforcement learning ($B^\textrm{DRL}$)
		
		\State Run a short simulation and collect the local state $s^l$ and the corresponding action of conventional SGS model $a^\textrm{SL}$
		\State Store the datasets $\left\lbrace s^l, a^\textrm{SL}\right\rbrace$ in $B^\textrm{SL}$
		\For{$t=1$ \textbf{to} $N^\textrm{SL}$}
		\State Sample mini-batch of $N$ datasets $\left\lbrace s^l, a^\textrm{SL}\right\rbrace$ from $B^\textrm{SL}$
		\State Update $\omega \leftarrow \mathrm{argmin}_{\omega} N^{-1} \sum (a^\textrm{SL} - \pi_\omega(s^l))^2$ 
		\EndFor		
		\State
		\For{$t=1$ \textbf{to} $N^\textrm{DRL}$}
		\State Perturb actor $\omega^\epsilon \leftarrow \omega + \epsilon$ with OU process
		\State Run LES with $\pi_{\omega^\epsilon}$ for $\Delta t^\textrm{DRL}$ and get the local and global state $s_t^l, s_t^g$ and the corresponding action $a_t$ $(=\pi_{\omega^\epsilon}(s_t^l))$ 
		\If{$t>N^s$}
		\State Calculate the reward $r_{t-N^s}$		
		\State Store the transitions $\left\lbrace s_{t-N^s}^l, s_{t-N^s}^g, a_{t-N^s}, r_{t-N^s} \right\rbrace$ in $B^\textrm{DRL}$		
		\State Sample mini-batch of $N$ datasets $\left\lbrace s^l, s^g, a, r\right\rbrace$ from $B^\textrm{DRL}$
		\State Update $\theta \leftarrow \mathrm{argmin}_{\theta} N^{-1} \sum (r - Q_\theta(s^l,s^g,a))^2$ 
		\State Update $\omega \leftarrow \mathrm{argmax}_{\omega} N^{-1} \sum Q_\theta(s^l,s^g,\pi_{\omega}(s^l))$ 	
		\EndIf	
		\EndFor	
	\end{algorithmic}
\end{algorithm}

The actor DNN, which is a mapping function between the local state $s_t^l$ and the local action $\tau_{ij}$, is a fully connected NN that consists of six hidden fully connected layers with 128 hidden units, and the output is six components of $\tau_{ij}$. The reason we chose a fully connected NN is that the ultimate goal in our application of DRL is to develop a universal pointwise SGS model which can be easily applied to a variety of turbulent flows. An introduction of a convolutional NN requires additional considerations of the structure of meshes such as size and type, which severely hinders generalizability of the developed SGS model. The critic DNN, which estimates the long-term reward based on total states $s_t^l$, $s_t^g$, and action $\tau_{ij}$, is a two-dimensional convolutional NN that efficiently considers homogeneous characteristics. When the $8\times1\times8$ (in the $x\times y\times z$) region is used as the input information, the critic DNN consists of six convolution layers, three average-pooling layers, and three fully connected layers. The block in which the $2 \times 2$ average pooling layer is applied after the two $3 \times 3$ convolution operations is repeated three times, and then it is connected to the fully connected layer. The number of hidden feature maps is $\{64, 128, 128, 256, 256, 256 \}$ for each convolution layer, and the number of hidden units for each fully connected layer is $\{256, 256\}$. The rectified linear unit with a slope of 2 is adopted for the nonlinear function that is applied after the convolution and full connection operations, except for the last layer. The trainable parameters in the actor and critic were initialized based on \citet{he2015deep}. The inputs of the actor are normalized in wall units such as $Re_\tau^{-1}\bar{\alpha}_{ij}$, because in the deep learning of turbulence problems, the universal predictability of wall-bounded turbulence insensitive to the Reynolds number was observed in wall-unit scaling of input and output variables \citep{Kim2020b,Kim2021,Park2021}, although it is not proven that subgrid-scale turbulence is universal in general. On the other hand, for a better prediction of the value function, we applied normalization techniques using the adaptive scaling values, which were calculated from the replay buffer data every 1000 DRL iterations. The reward is normalized by its root-mean-square value, and the inputs of the critic are normalized to have zero mean and a standard deviation of one.

To collect the datasets for the pre-training of the actor, a simulation for as short as tens of wall time units is sufficient, and the number of iterations for training ($N^\textrm{SL}$ in Algorithm \ref{Alg1}) is 50,000. Only a few minutes were consumed. To find an optimal policy, the number of iterations of DRL ($N^\textrm{DRL}$ in Algorithm \ref{Alg1}) is $10,000-15,000$. It takes approximately $2-4$ days using a single GPU machine (NVIDIA Titan Xp), predominantly depending on the mesh size of the LES and the calculation complexity of the actor DNN. The critic and actor networks were trained with a learning rate of $10^{-4}$ and $2.0\times 10^{-6}$, respectively. The mini-batch size was 64, and the Adam optimizer \citep{Kingma2014} was used. The size of the replay buffer was set to 64000, and the DRL algorithm was implemented using an open-source library {\it TensorFlow} \citep{Abadi2016}.

In LES for DRL, initial fields are obtained from a simulation with a 0-model ($\tau_{ij} = 0$), indicating that the initial states are very far from the target. The time interval of LES $\Delta t^\textrm{LES}$ was adjusted to maintain a Courant--Friedrichs--Lewy (CFL) coefficient of 0.15 and was approximately 0.18 in wall units for the case of Reynolds number $Re_\tau = 180$ and grid resolution $(\Delta x^+, \Delta z^+)=(70.7,35.3)$. For LES with a coarse resolution $(\Delta x^+, \Delta z^+)=(94.2,47.1)$ and LES with a higher Reynolds number $Re_\tau = 360$, the CFL coefficients were fixed as 0.1 and 0.18 for similar time intervals in wall units, respectively. The DRL time step $\Delta t^\textrm{DRL}$ between two consecutive states was 30 time steps of LES, which correspond to approximately 5.4 in wall time units. For DRL, only one episode of long LES was carried out because any time in a simulation can be considered as an initial field for another simulation, and the rapidly developing flow from the initial condition is not useful data for training. When the LES diverged, the LES inevitably restarted with the initial field. The length of the LES is approximately $1.1\times10^5$ ($=\Delta t^\textrm{DRL} \times N^\textrm{DRL}$) in wall time units. Furthermore, eight parallel LESs with different exploration noises of the actor were conducted to collect diverse data and shorten the training time, as \citet{Rabault2019a} showed the effectiveness of multi-environment DRL in controlling the flow around a cylinder. More details on DRL algorithms can be found in \citep{Kim2022}.

\section{Results}\label{sec3}

This section presents the results from investigations of the effects of hyperparameters, physical constraints, changes in training environments, and evaluations of the trained model in the trained and untrained flows. Through various training cases given in Table \ref{t.train}, the effectiveness and robustness of our DRL algorithm are investigated. First, we observe the existence of extreme instability in the DRL of turbulent flow and then verify the effect of the direct accumulation of future statistical results and the effect of physical constraints. Then, it is confirmed that DRL is capable of training SGS models with diverse targets, such as statistics for different grid resolutions and different Reynolds numbers. Next, we evaluated the trained SGS models in both trained and untrained environments and performed a quantitative comparative analysis with conventional SGS models. Finally, we compared the SGS stresses of the trained model with those of the fDNS and conventional models and present a possible reason for the limitation of the linear eddy-viscosity models.

\subsection{Effect of hyperparameters in DRL}

In general, reinforcement learning algorithms are sensitive to the hyperparameters involved, and thus, appropriate choices are essential for successful learning. Through an observation of their effects, we present guidance for hyperparameter tuning. For a quantitative comparison, the accuracy of statistics with the weighted time-averaging operation is defined as 
\begin{equation}
	A_i = {1\over N_y} \sum_{j=1}^{N_y}-\left|S_i^\textrm{fDNS}(y_j)- S_i^\textrm{LES*}(y_j) \right|^{1/2} .
\end{equation}
Here, $A_i$ is the accuracy of statistics, including $\left<d\bar{u}/dy\right>$, $\left<\bar{u}'\bar{v}'\right>$, $\left<\tau_{xy}\right>$, and mean spanwise shear stress $\left<d\bar{w}/dy\right>$ in wall units. $S_i^\textrm{LES*}$ represents the weighted time-averaged statistics using Eq. (\ref{eq24}). The reward used for actual training is directly related to the sum of $A_{\left<d\bar{u}/dy\right>}$ and $0.5A_{\left<\bar{u}'\bar{v}'\right>}$. $A_{\left<\tau_{xy}\right>}$ and $A_{\left<d\bar{w}/dy\right>}$ are not directly considered in the training. $A_i$ converges at a negative value, even in the optimal model, because the time-averaged length is short.

\begin{figure*}
	\centerline{\includegraphics[width=1.8\columnwidth]{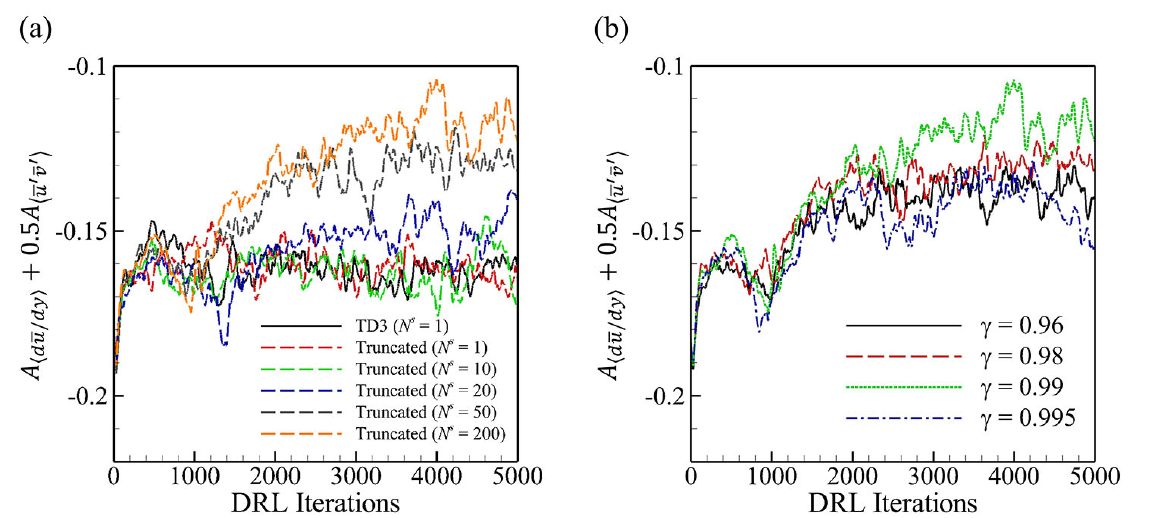}}
	\caption{Effect of reward accumulation on reinforcement learning of turbulent flows.}
	\label{fig04}
\end{figure*}

The important hyperparameters in Eq. (\ref{eq24}) are $\Delta t^{DRL}, \gamma$, and $N^s$, which are related to the time scale of learning. As explained in Section \ref{sec2.3}, $\Delta t^{DRL}$ was selected to be 5.4 wall time units, which is the shortest time scale guaranteeing the diversity of input data. Various values of $N^s$ and $\gamma$ were tested for the given $\Delta t^{DRL}$.
First, we demonstrate the cumulative effect of the reward by changing $N^s$ for $\gamma$ fixed at 0.99. We compared five DRLs, including the case of using instantaneous reward in Eq. (\ref{eq24}) (truncated (simplified) algorithm with $N^s = 1$), the case of using an indirect cumulative reward by the Bellman Eq. (\ref{eq7}) (TD3 algorithm with $N^s = 1$), and the cases of using a direct cumulative reward by Eq. (\ref{eq24}) (truncated (simplified) algorithm with $N^s = 10, 20, 50, 200$). The following values were used: a Reynolds number of $Re_\tau=180$, number of grids $(N_x, N_y, N_z)=(32,49,32)$ after dealising, grid resolution of $(\Delta x^+,\Delta z^+)=(70.7,35.3)$ in wall units, target statistics of $\left<d\bar{u}/dy\right>$ and $\left<\bar{u}'\bar{v}'\right>$, and the output of the actor DNN has six components of SGS stresses. Their accuracy results, which were calculated by averaging over 200 DRL steps, are shown in Fig. \ref{fig04}(a). Because of insufficient data in the replay buffer, some degradation from 0 to 1000 DRL iterations is usually observed. After the initial period, however, the performance according to the policy iterations became highly different. The instantaneous reward ($N^s=1$) did not provide an adequate direction for the policy update, therefore the accuracy did not increase. Similar results were observed even with the use of the Bellman equation. On the other hand, when using the reward calculated by directly accumulating future evolution, the statistical accuracy steadily increased. At $N^s = 20$, the reward starts improving. It appears that when $N^s$ is 50 or higher, the performance becomes saturated. This can be understood by the fact that for a given $\gamma=0.99$, the time scale of the decaying weight in the accumulation of reward can be estimated by $-\Delta t^{DRL}/\ln \gamma \approx 100\Delta t^{DRL}$ and that for $N^s \geq 200$, the accumulation has no further effect. Furthermore, $N^s=50$ corresponds to 250 wall time units, suggesting that the reaction time scale is below 250.

We also tested different values of $\gamma$ for fixed $N^s=200$, as shown in Fig. \ref{fig04}(b), clearly indicating that the performance for $\gamma=0.99$ is optimal. This suggests that there exists an optimal reward time scale in the weight decay of reward accumulation, which is $100 \Delta t^{DRL}$ or approximately 500 wall time units. This time scale may be related to the physical time scale of turbulence. In DNS, the life-time scale, which is the lifetime of turbulent structures, of the streamwise wall-shear stress is approximately 60 in wall units \citep{Quadrio2003}, indicating that the relevant statistics can fluctuate significantly within the reward time scale. Thus, it is reasonable to use the returns over the time period that is of the order of 10 times larger than the life-time scale. Strictly speaking, the optimal time scale is more relevant to the change in state than the change in action. Even in the same channel flow, the optimal time scale might differ across problems, depending on the information used to define the reward function. Proper definition of the reaction time scale is not easy, because some time scales can be changed depending on the method of perturbation of action. For an appropriate choice of the time scale, relevant future work is needed.

The above test results indicate that the long-term response is essential information for training, because the flow reaction for a short time scale ($\Delta t^\textrm{DRL} \approx 5$ in wall time units) is poorly correlated with the corresponding SGS stress. However, the function approximation of the long-term reward based on the Bellman equation is inaccurate in turbulent flows, and the direct accumulation of returns during the $N^s \ge 1$ period is decisive for successful learning. Although the training might be improved through the application of other DRL algorithms, we used our simplified algorithm because of its satisfactory performance. 

\begin{figure*}
	\centerline{\includegraphics[width=1.8\columnwidth]{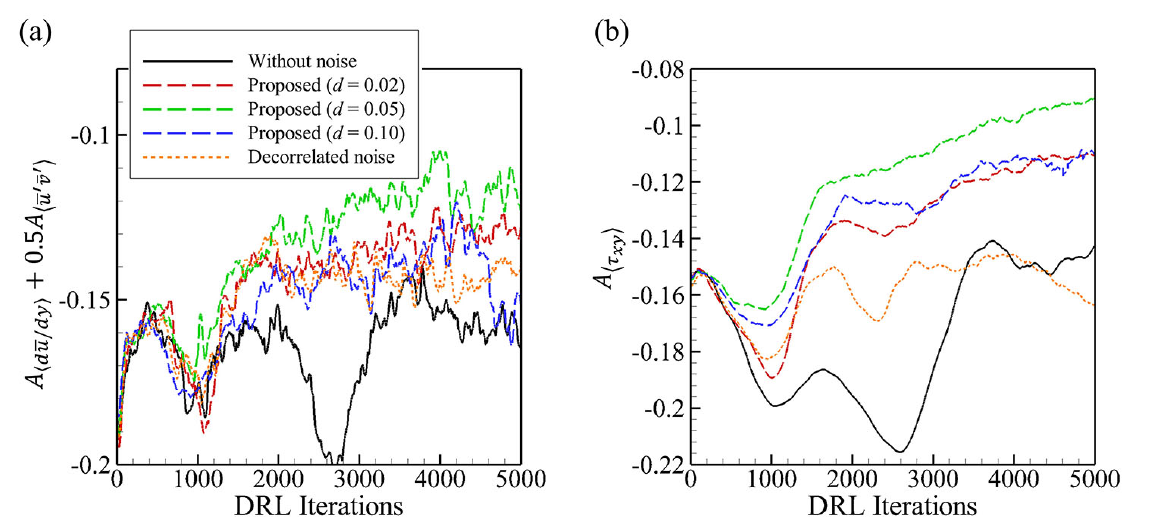}}
	\caption{Effect of exploration on reinforcement learning: (a) the sum of the accuracies of $\left<d\bar{u}/dy\right>$ and $\left<\bar{u}'\bar{v}'\right>$; (b) the accuracy of $\left<\tau_{xy}\right>$.}
	\label{fig05}
\end{figure*}

However, exploration methods and noise strength critically affect DRL performance. Under the same training environments as the above tests, three types of exploration including zero noise, (proposed) parameter space noise with the OU process, and (decorrelated) Gaussian noise were considered, and the results of statistical accuracy are given in Fig. \ref{fig05}. The case without noise showed strong oscillations in $A_{\left<d\bar{u}/dy\right>}+0.5A_{\left<\bar{u}'\bar{v}'\right>}$, because its exploration depends only on the training of the DNN, and the correlation of collected data in the replay buffer is too high. The decorrelated noise, with a $20\%$ intensity of the root-mean-square (rms) of instantaneous action fluctuations in each wall-normal location, could prevent the oscillation, but the increase in $A_{\left<\tau_{xy}\right>}$ was hardly observed, and the performance was almost similar to that of the case without noise. We also verified that the change in noise intensity had no significant effect. However, with the proposed exploration, a critical improvement in DRL performance is observed. We tested the effect of noise intensity $d$ ($=1-{1 \over K}\sum_k R_k$ in Section \ref{sec2-4}) through the cases of $d=$ 0.02 (weak), 0.05 (moderate), and 0.1 (strong). Weak noise resulted in relatively slow training performance, and moderate noise showed the fastest and most stable learning results. When strong noise was used, a slight oscillation was observed. This is because it frequently generated a bad action that caused the LES to diverge, preventing it from reaching the optimal state. As observed in Fig. \ref{fig05}(a), a steep decrease in accuracy in the case of strong noise does not imply performance degradation of the actor DNN and is only relevant to the initialization of diverged LES. In short, we found that the proposed exploration method is very effective in the DRL of turbulent flows, and finding the optimal strength $d$ of noise is important for cost reduction. Although the optimal $d$ might be different for each problem, we carried out DRLs using the same $d$ of 0.05 for the remaining cases of different LES environments and confirmed that DRLs can find the optimal DNN for all cases.

Finally, we tested the effect of the number of hidden layers ($N_h$) in the actor DNN, which is an important hyperparameter. For $N_h=3$, the DRL cannot find an optimal model in a coarse LES. This is probably because a relatively shallow NN has difficulty in representing strong nonlinearity between input and optimal actions and the exploration by shallow ones is less diverse. For $N_h = 6$ and 9, DRL could find optimal SGS models successfully, although their training speeds are slightly different. Both deep models provided good performance, but the deeper model required a higher computational cost to produce the SGS stresses for LES. Therefore, for the remaining cases, we used a DNN composed of six hidden layers.

\subsection{Effect of physical constraints on deep reinforcement learning}

The method of applying physical constraints, including the reflectional equivariance and the boundary condition to DRL, is provided in Section \ref{sec2-5}. We tested their effects for four cases: DRL with no physical constraint ($\textrm{DRL180}^{pure}$), DRL with the boundary condition constraint ($\textrm{DRL180}^b$), DRL with the reflection equivariance constraint in the wall-normal and spanwise directions ($\textrm{DRL180}^r$), and DRL with both constraints ($\textrm{DRL180}^{b,r}$, which is interchangeably denoted as $\textrm{DRL180}$). In all cases, the Reynolds number $Re_\tau$ is 180, the number of grids $(N_x, N_y, N_z)$ after dealising is $(32,49,32)$, the grid resolution $(\Delta x^+,\Delta z^+)$ is $(70.7,35.3)$ in wall units, the target statistics are $\left<d\bar{u}/dy\right>$ and $\left<\bar{u}'\bar{v}'\right>$, and the actor DNN produces six components of $\tau_{ij}$ from the local state information. The same hyperparameters were used for all cases. 

\begin{figure*}
	\centerline{\includegraphics[width=1.8\columnwidth]{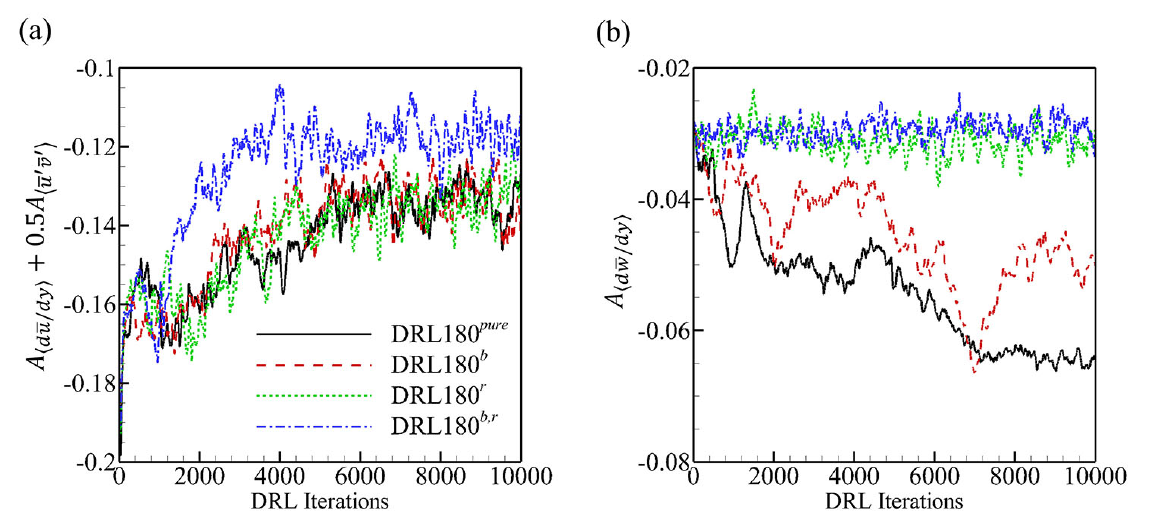}}
	\caption{Effect of physical constraints on reinforcement learning. $\textrm{DRL180}^{pure}$, $\textrm{DRL180}^{b}$, $\textrm{DRL180}^{r}$, and $\textrm{DRL180}^{b,r}$ have no physical constraints, boundary constraints, reflectional equivariance constraints, and both physical constraints, respectively. (a)  the accuracy of $\left<d\bar{u}/dy\right>$ and $\left<\bar{u}'\bar{v}'\right>$. (b) the accuracy of $\left<d\bar{w}/dy\right>$ related to the generation of nonphysical mean velocity in the spanwise direction.}
	\label{fig06}
\end{figure*}

The behavior of $A_{\left<d\bar{u}/dy\right>}+0.5A_{\left<\bar{u}'\bar{v}'\right>}$ and $A_{\left<d\bar{w}/dy\right>}$ with DRL iterations is shown in Fig. \ref{fig06}(a,b). First, regarding the boundary condition effect, Fig. \ref{fig06}(a) clearly shows that the learning of $\textrm{DRL180}^{b,r}$ is faster and more stable than $\textrm{DRL180}^{pure}$. $\textrm{DRL180}^{b}$ is only slightly better than $\textrm{DRL180}^{pure}$, because the grid size considered for model construction has a similar effect in the wall-resolved mesh. We found that when using the model without considering the grid size, the boundary constraint can yield a more significant effect (not shown here). In $\textrm{DRL180}^{pure}$, the resolved velocity could be suddenly changed by the nonphysically strong SGS mean shear stress near the wall during the exploration and training process, even though the reward function provided an indirect guide for the wall boundary condition. This indicates that it can be much more effective to fundamentally remove nonphysical elements in the construction of a DNN than to provide a guide through the reward function.

The effect of the reflectional equivariance constraint is notably observed in $A_{\left<d\bar{w}/dy\right>}$, as shown in Fig. \ref{fig06}(b). Because the spanwise statistics were not used in the reward function, the maximization of $A_{\left<d\bar{w}/dy\right>}$ by $\textrm{DRL180}^{pure}$ was not guaranteed, resulting in the generation of spanwise mean velocity. However, those of $\textrm{DRL180}^{r}$ and $\textrm{DRL180}^{b,r}$ were stationary regardless of training, although the value changes according to the time-average length. It is possible to provide a guide for $A_{\left<d\bar{w}/dy\right>}$ by changing the reward function, but it is imperfect and cannot reflect the equivariance property. Thus, it is essential to directly impose the reflectional equivariance property on the DNN.

Overall, $\textrm{DRL180}^{b,r}$, which reflects all physical constraints (considered in the present work), is more robust and stable compared with other cases. The reflectional equivariance prevented the generation of statistically anti-symmetric flows in the wall-normal and spanwise directions, and the boundary constraint effectively suppressed the unphysically high mean SGS shear stress near the wall. Therefore, all other DRLs in Table \ref{t.train} were performed with both physical constraints.

\subsection{Actual LES in the same environment as the training one}\label{sec3-3}

\begin{table*}
	\begin{center}
		\begin{tabular}{cccccc}
			&Trained SGS model  & Characteristic & $Re_\tau$ & $(N_x, N_y, N_z)$ & $(\Delta x^+, \Delta z^+)$ \\ \hline 
			&$\textrm{DRL180}$ & no constraint  & 180 & (32,49,32) & (70.7,35.3)  \\[1mm]
			&$\textrm{DRL180(clip)}$ & zero backscatter & 180 & (32,49,32) & (70.7,35.3) \\[1 mm]
			&$\textrm{DRL180}(\nu_t)$ & eddy-viscosity assumption & 180 & (32,49,32) & (70.7,35.3) \\[1 mm]
			&$\textrm{DRL180c}$ & no constraint & 180 &  (24,49,24) & (94.2,47.1)  \\[1mm]			
			&$\textrm{DRL180c(clip)}$ & zero backscatter & 180 &  (24,49,24) & (94.2,47.1)  \\[1 mm]			
			&$\textrm{DRL360}$  & no constraint & 360 &(32,65,32) & (70.7,35.3) \\[1mm]					
		\end{tabular}
		\caption{Training environments. The effects of the model forms, grid resolution, and Reynolds number are investigated. The constraints of the reflectional equivariance and boundary conditions are applied to all cases, and the target statistics are $\left< d\bar{u} \over dy \right>$ and $\left< \bar{u}'\bar{v}' \right>$. The same domain size as that of the DNS is used. $N_x$ and $N_z$ are the number of Fourier modes in the homogeneous directions, and $N_y$ is the number of wall-normal grids. $\Delta x^+$ and $\Delta z^+$ are the grid sizes of the wall units. In all cases, the minimum size of the wall-normal grid is smaller than 0.8 wall units.}{\label{t.train}}
	\end{center}
\end{table*}

To investigate the applicability of DRL in different environments, we varied the Reynolds number and grid resolution in both the training and testing stages. In addition, we tested performance restrictions using the model form. The training environments are presented in Table \ref{t.train} for each trained SGS model. $\textrm{DRL180(clip)}$ and $\textrm{DRL180c(clip)}$ are models in which the SGS stress was explicitly clipped to zero when negative dissipation occured, and $\textrm{DRL180}(\nu_t)$ is a model using the linear eddy-viscosity form similar to the SVM and DSM and was also clipped for negative eddy viscosity. $\textrm{DRL180}$, $\textrm{DRL180(clip)}$, and $\textrm{DRL180}(\nu_t)$ were trained for 10,000 DRL iterations, whereas $\textrm{DRL180c}$, $\textrm{DRL180c(clip)}$, and $\textrm{DRL360}$ were trained for 15,000 DRL iterations to find the optimal solutions.

\begin{table}
	\begin{center}
		\begin{tabular}{ccccccc}
			&Testing case  & SGS model & $Re_\tau$ & $(L_x, L_z)$ & $(N_x, N_y, N_z)$ & $(\Delta x^+, \Delta z^+)$ \\ \hline
			&LES180  & $\textrm{DRL180}$ & 180 & $(4\pi, 2\pi)$ & (32,49,32) & (70.7,35.3) \\[1mm]		
			&  & $\textrm{DRL180(clip)}$ & - & - & - & - \\[1mm]		
			&  & $\textrm{DRL180}(\nu_t)$ & - & - & - & - \\[1mm]		
			&  & $\textrm{DRL360}$ & - & - & - & - \\[1mm]		
			&  & SVM & - & - & - & - \\[1mm]		
			&  & DSM & - & - & - & - \\[1mm]		
			&  & 0-model & - & - & - & - \\ \hline	
			&LES180c  & $\textrm{DRL180c}$ & 180 & $(4\pi, 2\pi)$ & (24,49,24) & (94.2,47.1) \\[1mm]
			&  & $\textrm{DRL180c(clip)}$ & - & - & - & - \\[1mm]		
			&  & $\textrm{DRL180}$ & - & - & - & - \\[1mm]
			&  & DSM & - & - & - & - \\[1mm]
			&  & 0-model & - & - & - & - \\ \hline
			&LES360  & $\textrm{DRL360}$ & 360 & $(2\pi, \pi)$ & (32,65,32) & (70.7,35.3) \\[1mm]
			&  & $\textrm{DRL180}$ & - & - & - & - \\[1mm]
			&  & DSM & - & - & - & - \\[1mm]
			&  & 0-model & - & - & - & - \\ \hline
			&LES720  & $\textrm{DRL180}$ & 720 & $(\pi, 0.5\pi)$ & (32,97,32) & (70.7,35.3) \\[1mm]
			&  & $\textrm{DRL360}$ & - & - & - & - \\[1mm]
			&  & DSM & - & - & - & - \\[1mm]
			&  & 0-model & - & - & - & - \\[1mm]
			
		\end{tabular}
		\caption{Testing environments for the trained SGS models and the conventional SGS models. The effect of change in model forms, grid resolution effect, and Reynolds number $Re_\tau$ are considered. In the $Re_\tau$ effect, the resolution in the homogeneous directions is the same as that of the training case in the wall units. In all cases, the minimum size of the wall-normal grid was smaller than 0.8 wall units.}{\label{t.test}}
	\end{center}
\end{table}

For a quantitative comparison, we conducted LESs with the trained SGS models, conventional SGS models, and no model (0-model). Testing cases, including the Reynolds number and grid resolution effects, are presented in Table \ref{t.test}. For each testing case, we have tested the trained SGS model from a random initial condition over a sufficiently long time for accurate assessment. First, the trained model was tested in the same environment as the training Reynolds number and grid size. We evaluated the trained model by observing various statistics of the resolved and unresolved flow variables that were time-averaged over a long time period.

\begin{figure*}
	\centerline{\includegraphics[width=1.8\columnwidth]{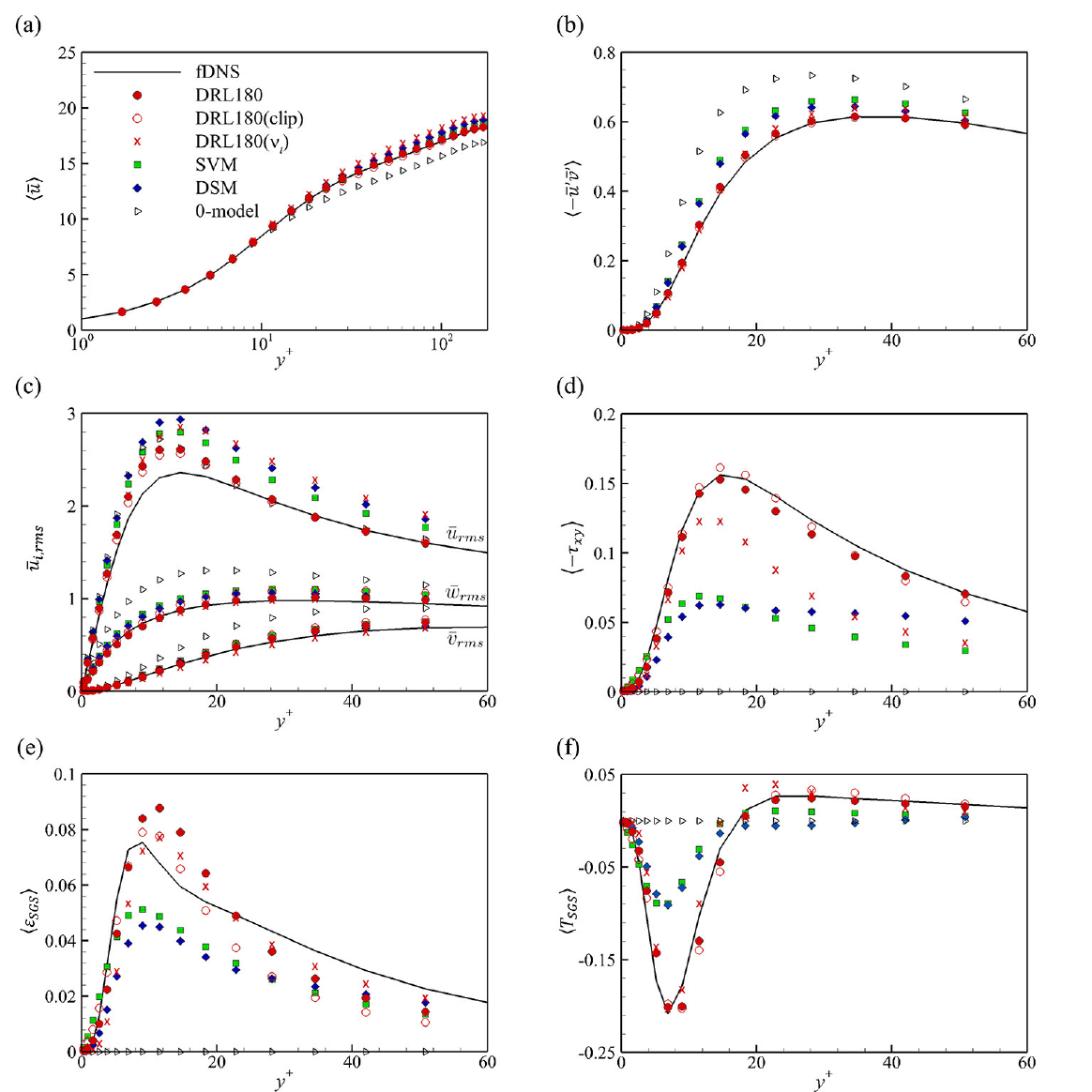}}
	\caption{Test results in training flow of $Re_\tau=180$ and $(\Delta x^+, \Delta z^+)=(70.7,35.3)$. The DRL models, including $\textrm{DRL180}$, $\textrm{DRL180(clip)}$, and $\textrm{DRL180}(\nu_t)$, were trained in the same environment. (a) and (b) are the streamwise mean velocity and the mean Reynolds shear stress, respectively, which are used for the DRL target. (c), (d), (e), and (f) are the root-mean-square (rms) of the velocity fluctuations, mean SGS shear stress, mean SGS dissipation, and mean SGS transport, respectively.}
	\label{fig07}
\end{figure*}

First, the statistics used as targets of DRL predicted by various SGS models are presented in Fig. \ref{fig07}(a,b). As expected from the results of statistical accuracy (Fig. \ref{fig06}), $\textrm{DRL180}$ produced accurate streamwise mean velocity $\left< \bar{u} \right>$, matching with that of fDNS (Fig. \ref{fig07}(a)). Interestingly, $\textrm{DRL180(clip)}$ embedded with a backscatter clipping operation showed similar results to $\textrm{DRL180}$. The SVM predicted $\left< \bar{u} \right>$ reasonably well through the adjustment of the model coefficient ($C_s=0.08$). On the other hand, $\textrm{DRL180}(\nu_t)$ and DSM overpredicted $\left< \bar{u} \right>$ in the buffer and logarithmic layers, and the 0-model highly underpredicted it. The Reynolds shear stress profile $\left< \bar{u}'\bar{v}' \right>$ is presented in Fig. \ref{fig07}(b). The SVM and DSM overestimated its magnitude, whereas the 0-model highly overestimated it because of insufficient SGS dissipation. In contrast, all the DRL models predictions were almost perfect. This result indicates that backscatter is not necessary for accurately predicting $\left< \bar{u} \right>$ and $\left< \bar{u}'\bar{v}' \right >$. Here, we briefly mention about the results of a standard supervised learning based on the filtered DNS data for training conducted by \citet{Park2021}. Among the various supervised learning models that they tested, the model using the same input information as ours did not predict $\left< \bar{u} \right>$ and $\left< \bar{u}'\bar{v}' \right >$ well. Furthermore, the actual LES results varied considerably depending on the input information, and some models numerically diverged. However, through the online learning process in which the temporal reaction of the SGS model in the actual LES is considered, the successful DRL guarantees the accuracy of the LES statistics used as the training target. It indicates that DRL using only statistical data for training can be superior to the standard supervised learning using expensive filtered DNS data. 

Next, we observe other statistics that are not directly considered in the reward function. The rms profiles of the velocity fluctuations $\bar{u}_{i,rms}$ are shown in Fig. \ref{fig07}(c). In $\bar{u}_{rms}$, the difference between models is most conspicuously observed. $\textrm{DRL180}$ and $\textrm{DRL180(clip)}$ predicted $\bar{u}_{rms}$ closer to fDNS, whereas the other models highly overestimated it. In addition, in $\bar{v}_{rms}$ and $\bar{w}_{rms}$, $\textrm{DRL180}$ and $\textrm{DRL180(clip)}$ showed slightly better results than the SVM and DSM and much better results than the 0-model. We presume that suppression of the fluctuations, $\bar{u}'$ and $\bar{v}'$, is an additional benefit in the process of making $\left< \bar{u}'\bar{v}' \right>$ precise. Although the prediction results varied slightly for each DRL case because the rms statistics were not directly considered as the reward, in all constraint-free DRL cases, we observed better prediction results than the other models.

Additionally, we presented statistics relevant to the SGS stress. The mean SGS shear stress $\left< \tau_{xy} \right>$ is the quantity that directly affects the viscous and Reynolds shear stresses by the mean balance Eq. (\ref{eq11}). Naturally, $\textrm{DRL180}$ and $\textrm{DRL180(clip)}$ produced accurate $\left< \tau_{xy} \right>$ well fitted to that of fDNS, as shown in Fig. \ref{fig07}(d). However, $\textrm{DRL180}(\nu_t)$ overpredicted $\left< \bar{u} \right>$, which underestimated the magnitude of $\left< \tau_{xy} \right>$ in the buffer and logarithmic layers, and SVM and DSM showed inaccuracy in $\left< \bar{u}'\bar{v}' \right>$, which yielded poor prediction of $\left< \tau_{xy} \right>$. We emphasize that the supervised learning model using the same input variable as $\textrm{DRL180}$ highly overpredicted the magnitude of $\left< \tau_{xy} \right>$ \citep{Park2021}.

Statistics relevant to the energy transfer of SGS stress, including the mean SGS dissipation and SGS transport, were investigated. SGS dissipation $\varepsilon_{SGS}=-\tau_{ij}\bar{S}_{ij}/{Re_\tau}$ means that the forward energy transfer from the resolved scales to the unresolved scales is important for predicting turbulence intensity well. In Fig. \ref{fig07}(e), the SVM and DSM produced insufficient mean SGS dissipation $\left< \varepsilon_{SGS} \right>$. The DRL models predicted $\left< \varepsilon_{SGS} \right>$ more closely to that of the fDNS, but a significant difference was observed. DRL models trained using the same settings produced different values of $\left< \varepsilon_{SGS} \right>$, and thus $\bar{u}_{i,rms} $ changed slightly for each DRL model (not shown here). In other words, in the model trained with only $\left< \bar{u} \right>$ and $\left< \bar{u}'\bar{v}' \right>$ as the DRL target, $\left< \varepsilon_{SGS} \right>$ was not uniquely determined.

Backward energy transfer from the unresolved scales to the resolved scales, the so-called backscatter, is a non-negligible quantity in fDNS and is often modeled for physical consistency. The scale-similarity model and DNN-based SGS models using more input information for better predictions are examples. However, inaccurate modeling of the backscatter can cause LES to diverge \citep{Guan2021}. For this reason, the DSM eliminates backscatter using the average operation in the homogeneous direction. In our study, backscatter was not observed prominently in $\textrm{DRL180}$, and the prediction performance of $\textrm{DRL180(clip)}$ is almost the same as that of $\textrm{DRL180}$. This means that considering backscatter is not a necessary condition for accurately predicting at least the streamwise mean velocity and mean Reynolds shear stress. It was reported that the backscatter is highly correlated with the strong Reynolds shear stress, which is relevant to near-wall events \citep{Kim1987,Piomelli1991,Piomelli1996}. To accurately reflect such physical phenomena, learning considering more diverse statistics, such as skewness and temporal correlation, might be necessary, and then the backscatter might be properly produced. A quest for which statistical quantity is needed to reflect the backscatter would be an interesting future work.

SGS transport $T_{SGS}=Re_\tau^{-1}{\partial (\tau_{ij} \bar{u}_{i}) / \partial x_j }$ is also an important result, as \citet{Voelker2002} reported that mean SGS transport $\left< T_{SGS} \right>$ is well predicted in optimal LES. As shown in Fig. \ref{fig07}(f), $\textrm{DRL180}$ and $\textrm{DRL180(clip)}$ predicted $\left< T_{SGS} \right>$ of the fDNS very well, whereas the SVM and DSM underpredicted its magnitude significantly, and $\textrm{DRL180}(\nu_t)$ showed a large error near $y^+=20$. Indeed, a common characteristic was observed in independently trained constraint-free DRLs with random initialization of trainable parameters. Consequently, to predict $\left< \bar{u} \right>$ and $\left< \bar{u}'\bar{v}' \right>$ well, accurate prediction of $\left< T_{SGS} \right>$ appears to be a necessary condition.

\begin{figure*}
	\centerline{\includegraphics[width=1.8\columnwidth]{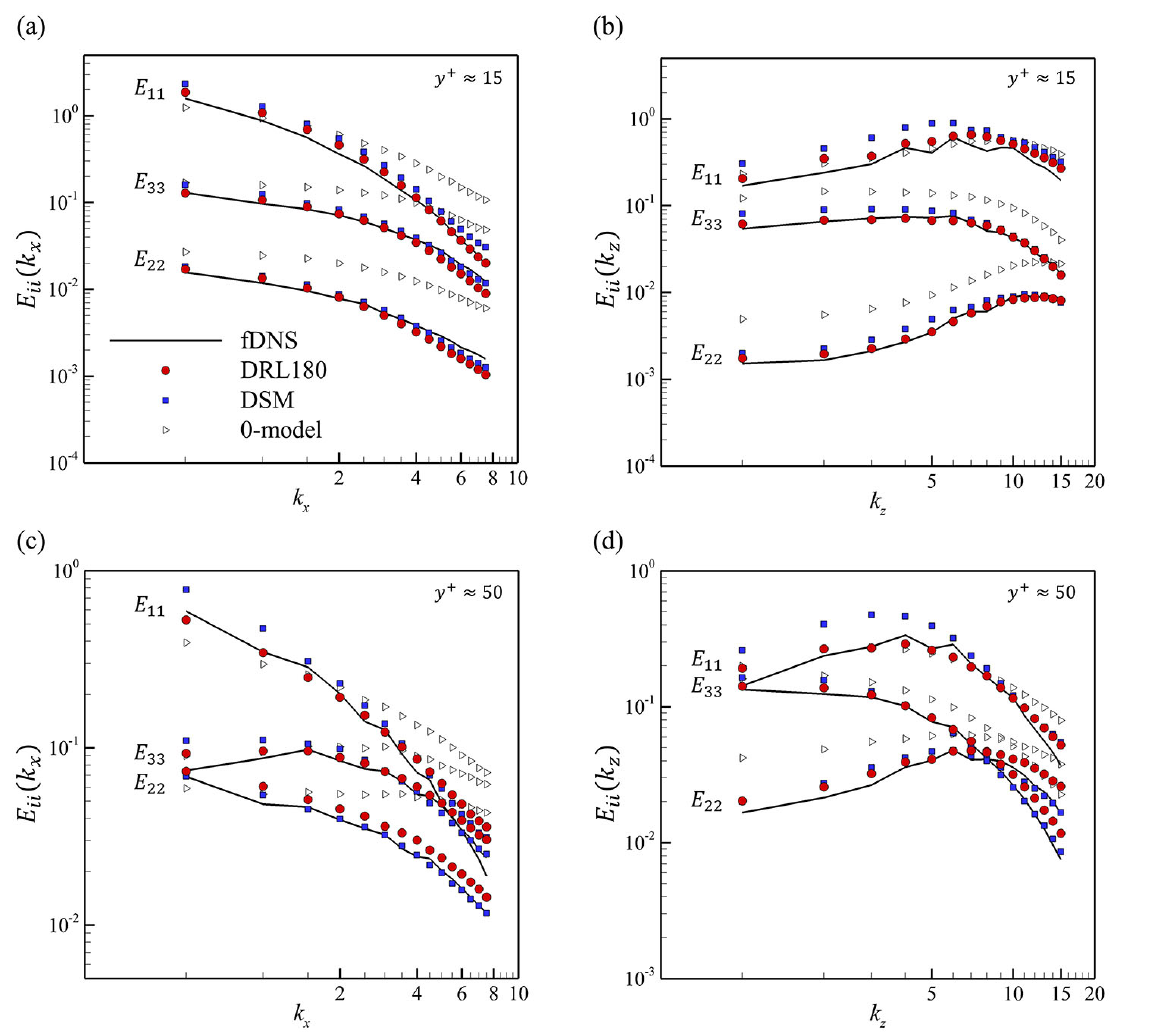}}
	\caption{One-dimensional energy spectra for $Re_\tau=180$. (a), (b), (c), and (d) are streamwise spectra at $y^+\approx15$, spanwise spectra at $y^+\approx15$, stream spectra at $y^+\approx50$, and spanwise spectra at $y^+\approx50$, respectively.}
	\label{fig08}
\end{figure*}

The one-dimensional energy spectra predicted by the DRL180, DSM, and 0-model are compared in Fig. \ref{fig08}. The streamwise energy spectra $E_{ii}(k_x)$ is defined as $\left< \hat{u_i}(k_x)\hat{u_i}^*(k_x)\right>$, where $\hat{u_i}$ and $k_x$ are the Fourier coefficients of the $\bar{u_i}$ and streamwise wavenumbers, respectively, and the superscript $^*$ denotes the complex conjugate. Spanwise, $E_{ii}(k_z)$ is similarly defined as $\left< \hat{u_i}(k_z)\hat{u_i}^*(k_z)\right>$. In the 0-model, significant overprediction of energy at high wavenumbers regardless of wall-normal height stands out because of the absence of the SGS effect. On the other hand, both \textrm{DRL180} and DSM accurately predict the energy at high wavenumbers, although they show some slight deviations. The error could be reduced by considering the rms of velocity fluctuations in the reward design. The difference between \textrm{DRL180} and DSM is observed at low streamwise and spanwise wavenumbers of the streamwise velocity. Unlike the overprediction of the DSM, \textrm{DRL180} shows reasonable results overall. This kind of performance of DRL180 is good evidence supporting that the DRL framework, which requires good statistical prediction, can indeed work if properly designed.

\begin{figure*}
	\centerline{\includegraphics[width=1.8\columnwidth]{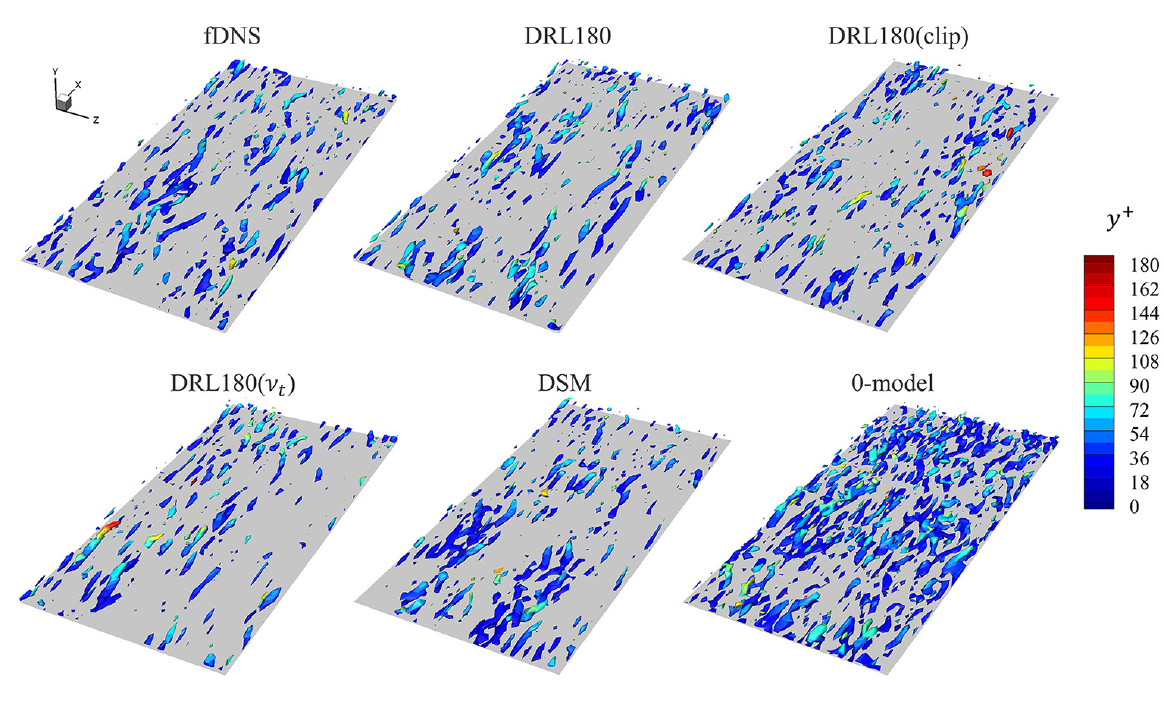}}
	\caption{Vortex visualization with $\lambda_2^+ = -0.005$. Flow is for $Re_\tau=180$ and $(\Delta x^+, \Delta z^+)=(70.7,35.3)$.}
	\label{fig09}
\end{figure*}

For the qualitative comparison of turbulent structures, the visualization of vortices using the $\lambda_2$ method was performed \citep{Jeong1995}. In Fig. \ref{fig09}, structures similar to the vortices of fDNS were observed in all of the DRL models. This might be due to an accurate prediction of the mean Reynolds shear stress directly related to the wall-normal motions. On the other hand, in the DSM, slightly more vortices were observed near the wall because of the insufficient SGS dissipation, and in the 0-model, too many non-physical structures were generated by the absence of SGS dissipation. Accurate prediction of the mean Reynolds shear stress appears to be important for representing near-wall turbulence structures.

In short, we confirmed that the LES with $\textrm{DRL180}$ and $\textrm{DRL180(clip)}$ predicted the statistics of both resolved and modeled scales much better than $\textrm{DRL180}(\nu_t)$ and conventional SGS models including the SVM, DSM, and 0-model. We found that accurate representations of the mean SGS shear stress and mean SGS transport are necessary to accurately predict the streamwise mean velocity and mean Reynolds shear stress, but the accuracy of mean SGS dissipation and backscatter are not essential requirements. Accurate prediction of SGS energy transfer might be relevant to high-order statistics or temporal behavior of the resolved variables.

\begin{figure*}
	\centerline{\includegraphics[width=1.8\columnwidth]{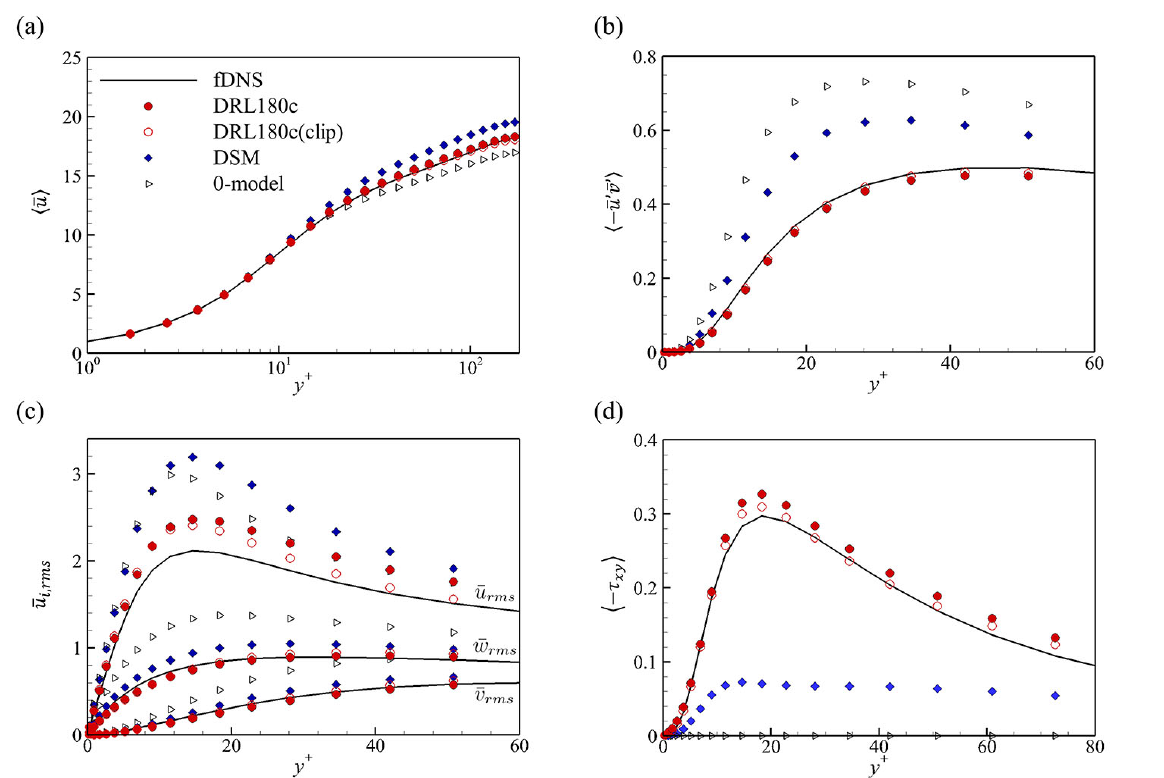}}
	\caption{Test results in training flow of $Re_\tau = 180$ and coarse resolution $(\Delta x^+, \Delta z^+)=(94.2,47.1)$. $\textrm{DRL180c}$ and $\textrm{DRL180c(clip)}$ were trained in the same environment. (a) and (b) are the streamwise mean velocity and mean Reynolds shear stress, respectively, used for the training target. (c) and (d) are the rms of the velocity fluctuations and the mean SGS shear stress, respectively.}
	\label{fig10}
\end{figure*}

Next, DRLs in harsher training environments, including the cases of coarse grid resolution ($\textrm{DRL180c}$ and $\textrm{DRL180c(clip)}$) and a higher Reynolds number ($\textrm{DRL360}$), were carried out. In fDNS, the maximum magnitude of the mean SGS shear stress in the coarse resolution $(\Delta x^+, \Delta z^+)=(94.2,47.1)$ was approximately twice as high as that in the fine resolution $(\Delta x^+, \Delta z^+)=(70.7,35.3)$. Most conventional SGS models highly underestimate the magnitude of the mean SGS shear stress, indicating that the degree of correction through learning is large. Nevertheless, we confirmed that the successful learning of $\textrm{DRL180c}$ and $\textrm{DRL180c(clip)}$ is possible using the same hyperparameters as those of $\textrm{DRL180}$, except for the number of DRL iterations. The statistical results of LES180c performed using the trained models, DSM, and the 0-model are provided in Fig. \ref{fig10}.

The streamwise mean velocity profiles predicted by $\textrm{DRL180c}$ and $\textrm{DRL180c(clip)}$ were consistent with those of fDNS, while significant overestimation and underestimation were observed in the DSM and 0-model, respectively (Fig. \ref{fig10}(a)). Similar results were also observed for the mean Reynolds shear stress, rms of velocity fluctuations, and mean SGS shear stress (Fig. \ref{fig10}(b,c,d), respectively). In the DSM and 0-model, the magnitude of the Reynolds stresses was generally overpredicted by insufficient SGS dissipation. In addition, the maximum magnitude of the mean SGS shear stress of the fDNS was three times larger than that of the DSM. On the other hand, we found that both DRL models can accurately predict these quantities. These results indicate that successful LES modeling is possible through DRL even in a situation where the effect of residual stress is dominant, and backscatter is also not essential for predicting the streamwise mean velocity and mean Reynolds shear stress.

\begin{figure*}
	\centerline{\includegraphics[width=1.8\columnwidth]{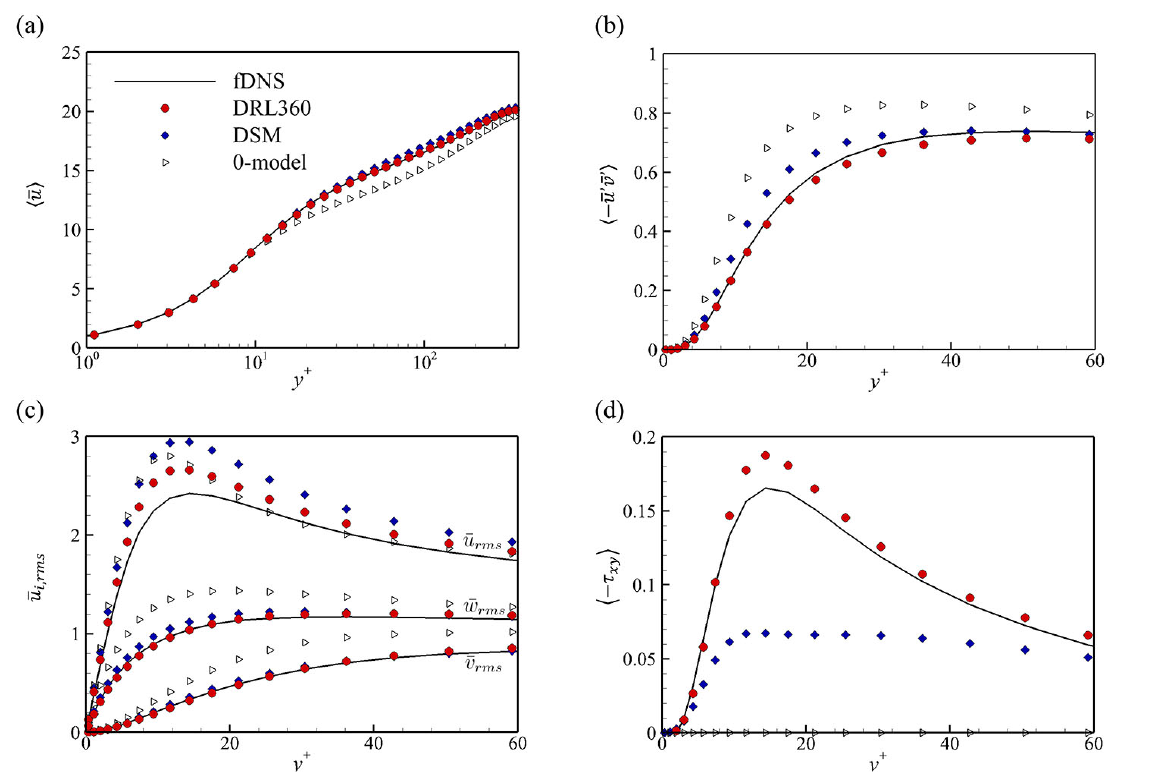}}
	\caption{Test results in training flow of a higher Reynolds number $Re_\tau=360$ and ($\Delta x^+$, $\Delta z^+$)=(70.7,35.3). The DRL model, $\textrm{DRL360}$, was trained in the same environment. (a) and (b) are the streamwise mean velocity and mean Reynolds shear stress, respectively, used for the training target. (c) and (d) are the rms of the velocity fluctuations and the mean SGS shear stress, respectively.}
	\label{fig11}
\end{figure*}

We also considered the DRL with LES with a higher Reynolds number. As the friction Reynolds number increases, the channel flow includes more varied physics in the wall-normal direction, such as the long logarithmic law region, which is associated with an increase in the action dimension. In addition, the relative proportion of the high mean SGS shear stress region is decreased compared to the case of $Re_\tau=180$; that is, the key information in learning exists sparsely. The difficulty of learning with higher $Re_\tau$ increases. Nevertheless, our DRL algorithm was able to find the optimal SGS model at $Re_\tau=360$ using the same hyperparameters as before. The results of the actual LES with the trained model ($\textrm{DRL360}$) are presented in Fig. \ref{fig11}. $\textrm{DRL360}$ can well predict all the given statistics, including the streamwise mean velocity, Reynolds stresses, and mean SGS shear stress, while LESs with the DSM and 0-model provide inaccurate statistical results, especially in the Reynolds stresses.

In summary, we found that the proposed DRL algorithm works robustly in various environments. As a result, we can obtain the DNN-based SGS model for accurately predicting target statistics, and as an additional advantage, the prediction accuracy of other statistics, such as velocity rms and mean SGS transport, was also improved compared to the conventional models. In addition, through changes in the model form, we found that using the eddy-viscosity form could cause large inaccuracies and that backscatter is not an essential factor in accurately predicting the streamwise mean velocity and Reynolds shear stress. Our results indicate that it is possible to develop a high-fidelity SGS model using only statistical information in real-world problems where DNS flow fields are not accessible. In the next subsection, we test the trained model on the LES in a new untrained environment.

\subsection{Test in untrained flows: higher Reynolds numbers and different grid resolution} 

The performance of the trained SGS models in unseen flows with different Reynolds numbers and grid resolutions is investigated in this section. Some features of turbulent channel flow are universal in wall units, and the most representative quantity is the mean velocity profile regardless of the Reynolds number, called the law of the wall (LOW). Recently, universality has also been observed in deep learning research, which finds a nonlinear relation between turbulence variables in an input-output framework. For example, \citet{Kim2021} showed that the deep learning model trained to reconstruct high-resolution turbulence from a low-resolution model at $Re_\tau=1000$ can also perform well at $Re_\tau=5200$ with wall unit scaling. This readily suggests that the relation between low-resolution (resolved scale) fields and SGS stress might be universal in channel flows with high Reynolds numbers. \citet{Park2021} reported that an NN-based SGS model trained using $Re_\tau = 180$ can produce reasonably accurate flow in the actual LES of $Re_\tau = 720$ in the same grid resolution in wall units. However, in the \textit{a priori} test, the model showed significant inaccuracy for $Re_\tau = 720$. It is still unclear whether the trained SGS model can be a universal function with respect to the Reynolds number. In DRL, there are non-deterministic parts in that it is not a direct supervised learning of the flow field, including physics, but consequently aims to match the target statistics. Therefore, we investigated the performance of DRL models for LESs with higher Reynolds numbers than trained models.

\begin{figure*}
	\centerline{\includegraphics[width=1.8\columnwidth]{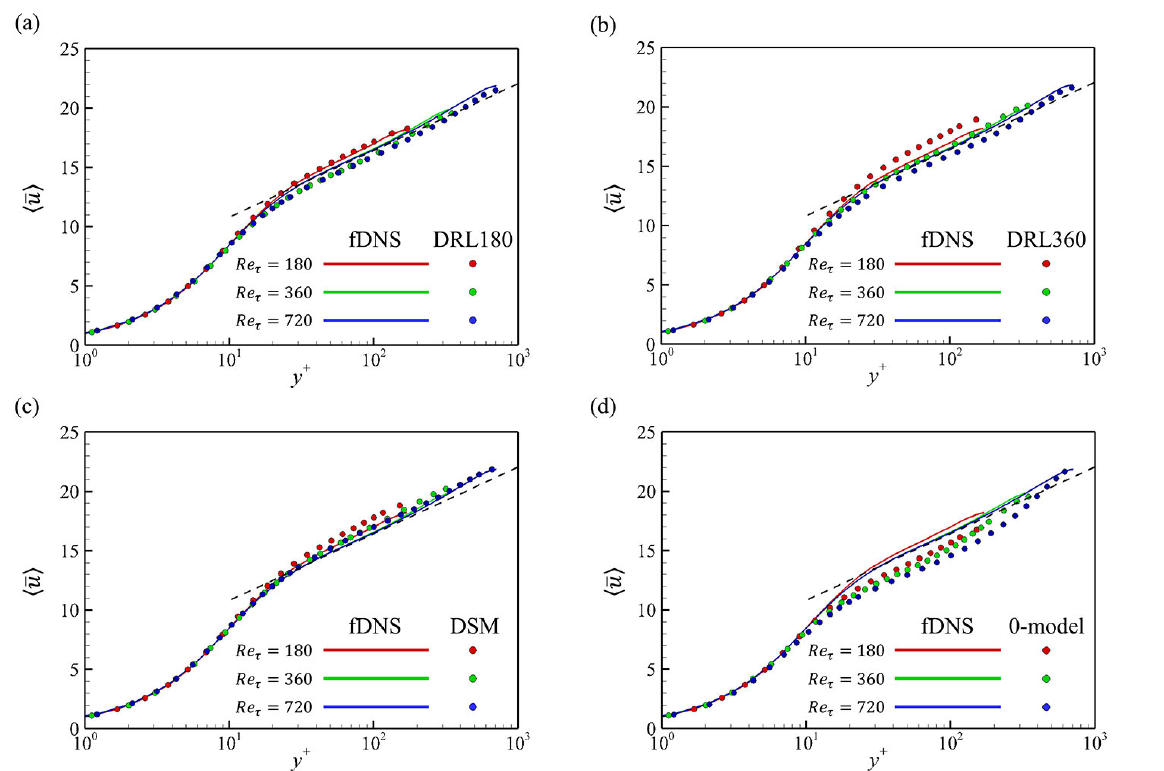}}
	\caption{Test results in flows of several Reynolds numbers $Re_\tau=180, 360, 720$ and $(\Delta x^+, \Delta z^+)=(70.7,35.3)$. (a) and (b) are the streamwise mean velocities of LESs with $\textrm{DRL180}$ trained at $Re_\tau=180$ and with $\textrm{DRL360}$ trained at $Re_\tau=360$, respectively. (c) and (d) are those of the LESs with DSM and the 0-model, respectively. Solid lines, symbols, and dashed lines denote fDNS, LES, and the law of the wall ($\left<\bar{u}\right>=0.41^{-1} \textrm{log}(y^+)+5.2$), respectively.}
	\label{fig12}
\end{figure*}

LESs of $Re_\tau=180, 360$, and 720 were performed using the $\textrm{DRL180}$ SGS model trained at $Re_\tau=180$, $\textrm{DRL360}$ trained at $Re_\tau=360$, DSM, and 0-model as listed in Table \ref{t.test}. The horizontal grid resolution in the wall units is the same as that in the training unit, and the input information of the actor DNN used to produce the SGS stress is normalized in the wall units of each flow. In the streamwise mean velocity profile in Fig. \ref{fig12}(a), unlike the perfect accuracy of $Re_\tau=180$, $\textrm{DRL180}$ slightly underestimated the slope in the buffer layer at $Re_\tau=360$. One possible reason for this mismatch is the low Reynolds number effect \citep{Moser1999}, in which the streamwise mean velocity profile of DNS at $Re_\tau=180$ slightly deviates from the LOW in the logarithmic region. As evidence supporting this, the results of $\textrm{DRL360}$ are presented in Fig. \ref{fig12}(b). $\textrm{DRL360}$ overestimated the mean velocity of $Re_\tau=180$. This indicates that the non-universal characteristic of $Re_\tau=180$ affects the generalization performance of the DRL model. Moreover, for $Re_\tau=720$, both DRL models slightly underestimated the slope in the buffer layer. This might be relevant to the grid resolution in the wall-normal direction. The characteristic that the accuracy differs according to the Reynolds number is also observed in the DSM and 0-model (Fig. \ref{fig12}(c,d)). The DSM slightly overestimated the mean velocity of all Reynolds numbers, whereas the 0-model highly underestimated it. In short, although the DRL models showed reasonable predictability of the mean velocity in unseen flows, DRL in an environment with a fixed simulation parameter is not sufficient for better generalization ability.

\begin{figure*}
	\centerline{\includegraphics[width=1.8\columnwidth]{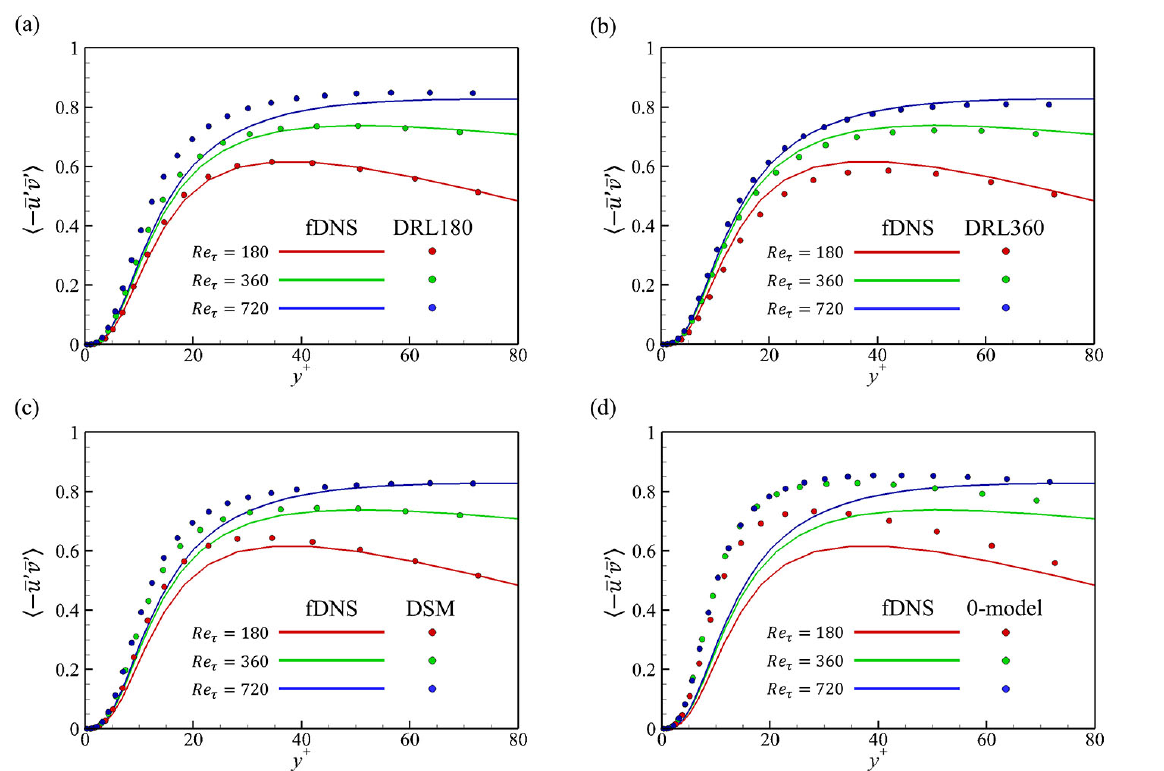}}
	\caption{Test results in flows of several Reynolds numbers $Re_\tau=180, 360, 720$ and $(\Delta x^+$, $\Delta z^+)=(70.7,35.3)$. (a) and (b) are the Reynolds shear stresses of LESs with DRL180 trained at $Re_\tau=180$ and DRL360 trained at $Re_\tau=360$, respectively. (c) and (d) are those of the LESs with DSM and the 0-model, respectively. All lines are fDNS results, and the symbols are LES results.}
	\label{fig13}
\end{figure*}

In the mean Reynolds shear stress profile, a different tendency is observed in Fig. \ref{fig13}. The LESs of $\textrm{DRL180}$ and $\textrm{DRL360}$ are quite precise in prediction even for untrained Reynolds numbers, while DSM overpredicted the mean Reynolds shear stress for all Reynolds numbers because of insufficient mean SGS shear stress, and the 0-model significantly overpredicted it. For other Reynolds stresses, the DRL models provided better predictions than the DSM and 0-model (not shown in this paper). Although DRL models appear to provide reasonable prediction accuracy, $\textrm{DRL180}$ shows an obvious tendency to overestimate the mean Reynolds shear stress as the Reynolds number increases. This indicates that the DRL trained using statistical data of a single Reynolds number might not capture universal elements for the mean velocity and Reynolds stress for various Reynolds numbers.

Finally, LESs were performed to evaluate the generalization of the grid resolution. In particular, we focus on the case of coarse grid resolution in homogeneous directions, for which the conventional SGS models usually work poorly, as shown in Fig. \ref{fig10}. As a result of performing coarse LES using $\textrm{DRL180}$, we found that it overpredicted the mean Reynolds shear stress and underpredicted the mean SGS shear stress (not shown here). This indicates that a model trained in a single environment might not perform well in a different environment. Such limitations have also been reported in classical supervised learning models. To develop a model that is generally applicable, DRL in multiple environments with different simulation parameters or DRL in a flow that includes more diverse physics seems to be essential.

\subsection{Characteristics of the trained model compared with linear eddy-viscosity models}

We confirmed that the trained model using the proposed DRL can accurately predict the statistics of the viscous and Reynolds shear stresses in the training environment. In this section, we analyze the characteristics of how the model can achieve such accuracy. For comparison, filtered DNS data and conventional linear eddy-viscosity models were used. However, we clearly state that having exactly the same characteristics as fDNS is not necessary for accurate LES. Furthermore, it is impossible to accurately predict the local SGS stress of fDNS from the local velocity gradient information owing to the input dependency, as \citet{Park2021} reported that the correlation coefficient between the SGS stress of fDNS and the DNN prediction based on the local velocity gradient is approximately 0.5. We focus on features commonly observed in fDNS and DRL models compared with linear-eddy viscosity models. The fDNS results were obtained by applying a cut-off filter in homogeneous directions to the DNS flow fields. To investigate how each SGS model produces residual stresses, all the SGS models were applied to the same resolved velocity gradient data of LES180 carried out with $\textrm{DRL180}$. This approach allows for a direct comparison between all tested SGS models.

\begin{figure*}
	\centerline{\includegraphics[width=1.8\columnwidth]{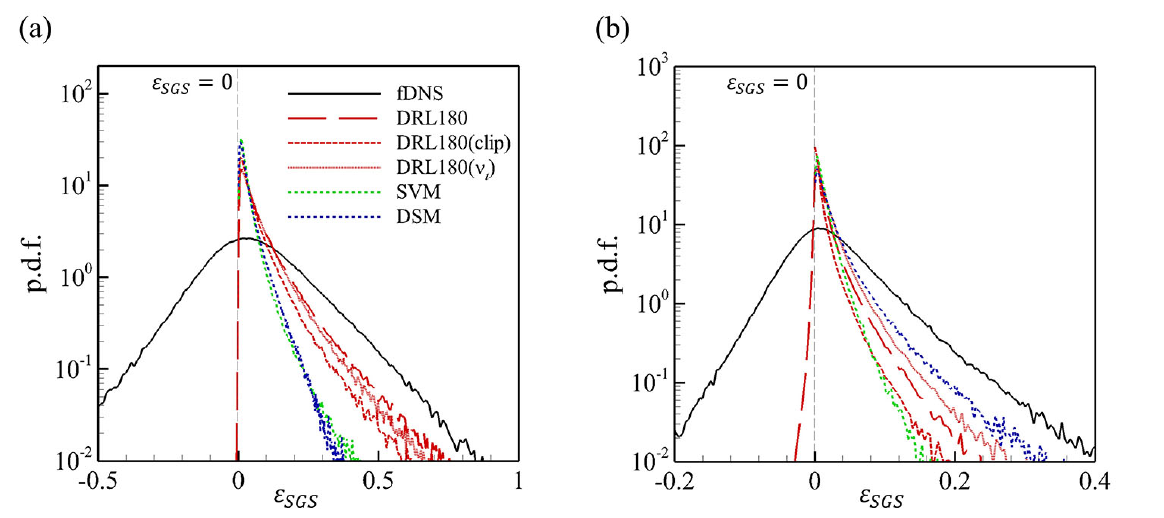}}
	\caption{Probability density functions (PDFs) of SGS dissipation. Results for $Re_\tau=180$ and $(\Delta x^+, \Delta z^+)=(70.7,35.3)$. (a) $y^+\approx 15$, (b) $y^+\approx 50$.}
	\label{fig14}
\end{figure*}

First, the probability density functions (PDFs) of SGS dissipation predicted by all the SGS models at two wall-normal locations are presented against that of fDNS in Fig. \ref{fig14}. Near the wall, relatively long positive tails are commonly observed in the SGS dissipation of the DRL models (Fig. \ref{fig14}(a)). The relatively high dissipation of $\textrm{DRL180}(\nu_t)$ compared with conventional SGS models resulted in the mean Reynolds shear stress being accurate; however, the restriction of eddy-viscosity from the streamwise mean velocity was highly overpredicted (Fig. \ref{fig07}). On the other hand, in $\textrm{DRL180}$ and $\textrm{DRL180(clip)}$ without the eddy-viscosity assumption, such distortion was not observed despite strong SGS dissipation. In Fig. \ref{fig14}(a), the considerably large negative dissipation observed in fDNS was not produced by all SGS models, indicating that the backscatter is not necessary for predicting viscous and Reyonlds shear stresses accurately. An interesting result is that at $y^+\approx 50$, weak backscatter is observed for $\textrm{DRL180}$ (Fig. \ref{fig14}(b)). Although backscatter is not an essential element, it can be generated owing to randomness in the training process, and it shows that backscatter does not necessarily cause numerical instability.

\begin{figure*}
	\centerline{\includegraphics[width=2.0\columnwidth]{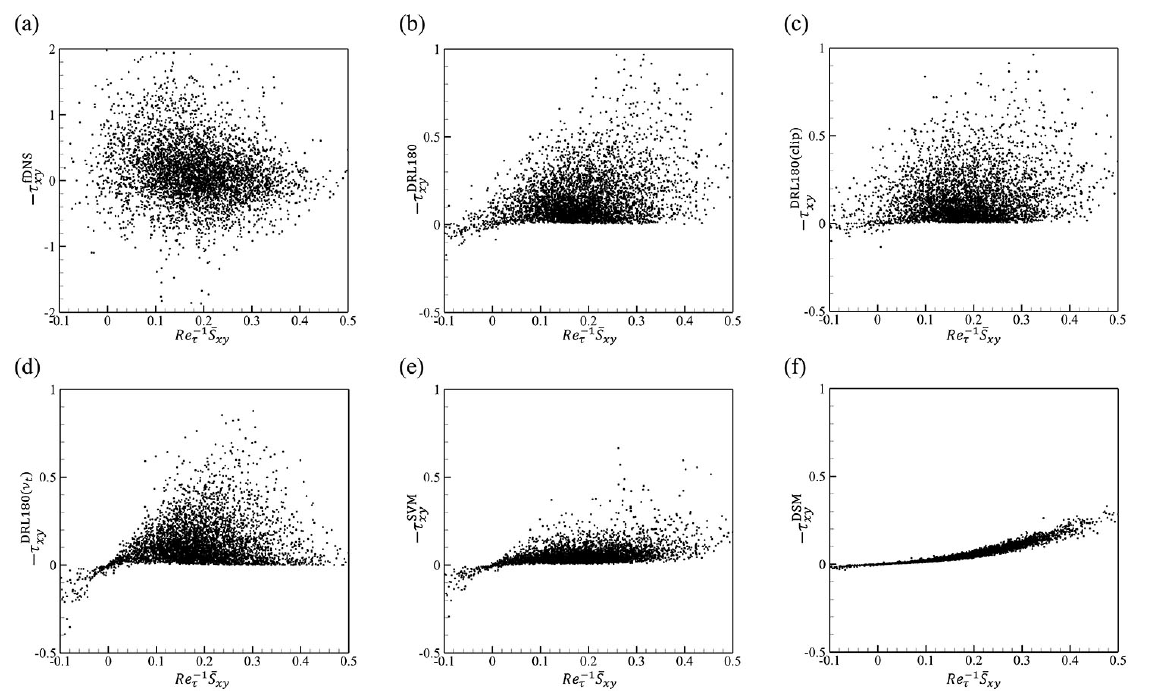}}
	\caption{Correlation between $\bar{S}_{xy}$ and $\tau_{xy}$ at $y^+\approx15$. Results for $Re_\tau=180$ and $(\Delta x^+, \Delta z^+)=(70.7,35.3)$. (a) fDNS, (b) $\textrm{DRL180}$, (c) $\textrm{DRL180(clip)}$, (d) $\textrm{DRL180}(\nu_t)$, (e) SVM, and (f) DSM.}
	\label{fig15}
\end{figure*}

\begin{figure*}
	\centerline{\includegraphics[width=2.0\columnwidth]{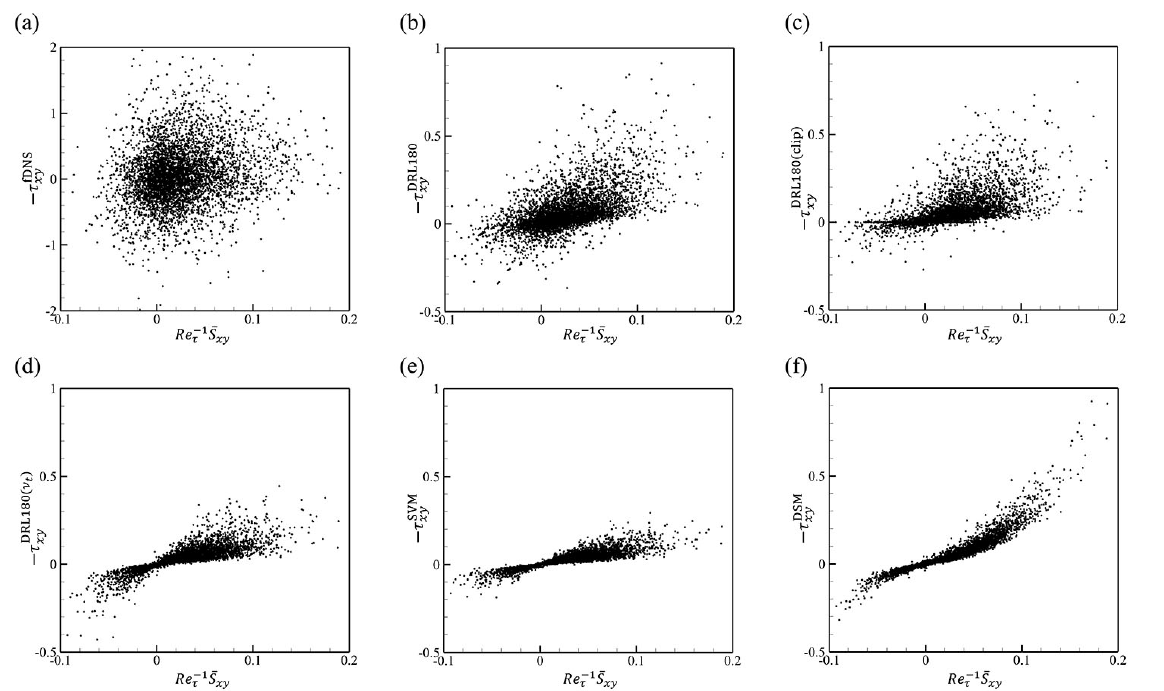}}
	\caption{Correlation between $\bar{S}_{xy}$ and $\tau_{xy}$ at $y^+\approx50$. Results for $Re_\tau=180$ and $(\Delta x^+, \Delta z^+)=(70.7,35.3)$. (a) fDNS, (b) $\textrm{DRL180}$, (c) $\textrm{DRL180(clip)}$, (d) $\textrm{DRL180}(\nu_t)$, (e) SVM, and (f) DSM.}
	\label{fig16}
\end{figure*}

\begin{table*}
	\begin{center}
		\begin{tabular}{cccccccccc}
			& Location  & Variables & fDNS & $\textrm{DRL180}$ & $\textrm{DRL180}^{DSM}$ & $\textrm{DRL180(clip)}$ & $\textrm{DRL180}(\nu_t)$ & SVM & DSM \\ \hline
			& $y^+\approx15$ & $\bar{S}_{xy}$, $-\tau_{xy}$ & -0.20 & \textbf{0.34} & \textbf{0.63} & \textbf{0.23} & 0.22 & 0.48 & 0.95 \\[1mm]
			& - & $|\bar{S}|$, $\varepsilon_{SGS}$ & 0.01 & 0.56 & 0.77 & 0.60  & 0.30 & 0.54 & 0.92 \\ \hline	
			& $y^+\approx50$ & $\bar{S}_{xy}$, $-\tau_{xy}$ & 0.19 & \textbf{0.58} & \textbf{0.94} & \textbf{0.53} & 0.82 & 0.87 & 0.94 \\[1mm]
			& - & $|\bar{S}|$, $\varepsilon_{SGS}$ & 0.28 & 0.73 & 0.80 & 0.71 & 0.67 & 0.78 & 0.88 \\[1mm]	
		\end{tabular}
		\caption{Correlation coefficient between the resolved and modeled flow variables. $\textrm{DRL180}^{DSM}$ was pretrained using the DSM, whereas other DRL models were pretrained using the SVM.}{\label{t.cor}}
	\end{center}
\end{table*}

Next, the correlation between $\bar{S}_{xy}$ and $-\tau_{xy}$ predicted by all the SGS models at $y^+\approx 15$ and 50 are presented in Fig. \ref{fig15} and \ref{fig16}, respectively. Interestingly, they are almost decorrelated in fDNS at both locations and are negatively correlated at $y^+\approx 15$. On the other hand, quadratic behavior is clearly observed in DSM, and a relatively high correlation is shown in SVM, although lower than DSM. In contrast, it is noteworthy that a fairly low correlation is observed in $\textrm{DRL180}$, $\textrm{DRL180(clip)}$, and $\textrm{DRL180}(\nu_t)$. At $y^+\approx 50$, a more significant difference is observed between the DRL models and eddy-viscosity models (Fig. \ref{fig16}). In the second and fourth quadrants, $\tau_{xy}$ of $\textrm{DRL180}$ and $\textrm{DRL180(clip)}$ can be produced regardless of the clipping operation, whereas it is not observed in $\textrm{DRL180}(\nu_t)$, SVM, and DSM because of the positively defined $\nu_t$. For quantitative comparison, Table \ref{t.cor} presents the correlation coefficient, which is defined as $R_{V_1V_2}=cov(V_1,V_2)/(\sigma_{V_1}\sigma_{V_2})$, and $V_i$, $cov$, and $\sigma$ are the variables, covariance, and standard deviation, respectively. In $R_{\bar{S}_{xy}\tau_{xy}}$, we can clearly confirm that the DRL models have a low correlation at $y^+\approx 15$ but relatively moderate correlation at $y^+\approx 50$. This indicates that the artificially high correlation between $\bar{S}_{xy}$ and $\tau_{xy}$ in the near-wall region excessively suppresses the fluctuation of $\bar{u}$, resulting in the degradation of the prediction accuracy. On the other hand, the DRL models pretrained using DSM showed a high correlation at both locations.

\begin{figure*}
	\centerline{\includegraphics[width=2.0\columnwidth]{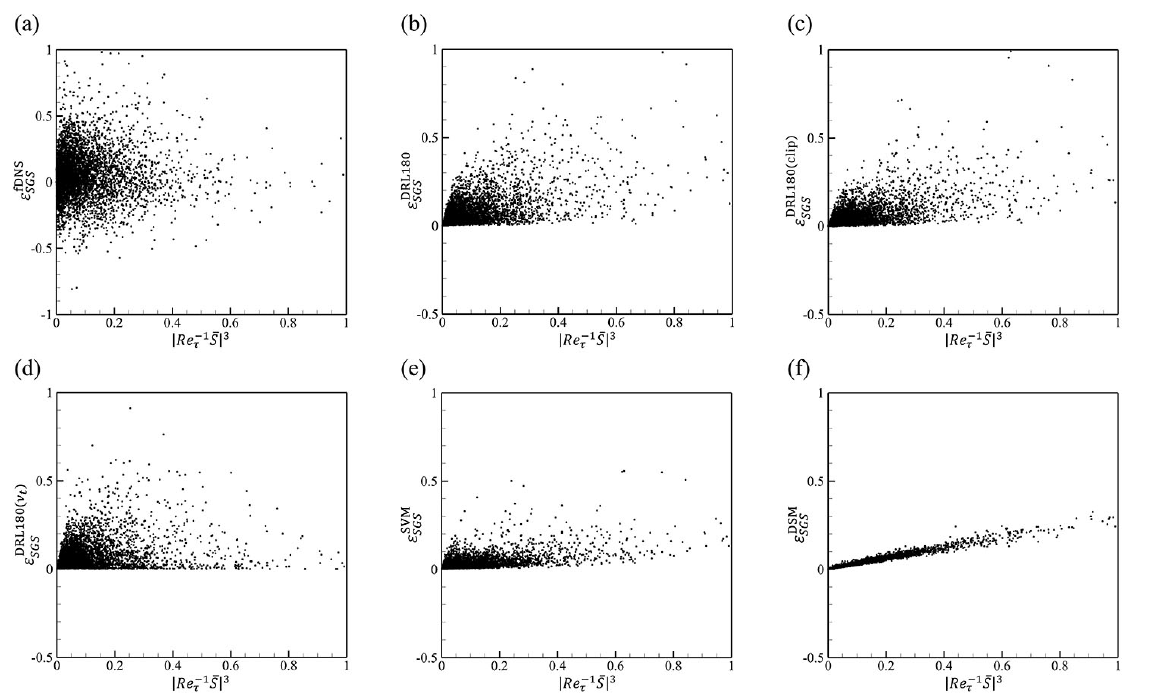}}
	\caption{Correlation between $|\bar{S}|^3$ and $\varepsilon_{SGS}$ at $y^+\approx15$. Results for $Re_\tau=180$ and $(\Delta x^+, \Delta z^+)=(70.7,35.3)$. (a) fDNS, (b) DRL180, (c) DRL180(clip), (d) DRL180($\nu_t$), (e) SVM, and (f) DSM.}
	\label{fig17}
\end{figure*}

Similarly, the correlation between $|\bar{S}|^3$ and $\varepsilon_{SGS}$ at $y^+\approx 15$ is presented in Fig. \ref{fig17}. By the relation ${Re_\tau}\varepsilon_{SGS}=-\tau_{ij}\bar{S}_{ij} = \nu_t |\bar{S}|^2= C^\textrm{DSM} |\bar{S}|^3$, a linear relationship is clearly observed in DSM, although its dynamic procedure produces weak fluctuations of the slope. On the other hand, such linearity is not observed in fDNS, with the correlation being close to zero (Table \ref{t.cor}). Other SGS models exhibit deviations from linearity. SVM shows a relatively weak deviation, whereas all of the DRL models show a relatively large deviation. Despite the constraints of the same eddy-viscosity form, $\textrm{DRL180}(\nu_t)$ produced a larger deviation than the SVM, indicating the existence of a high variance of $\nu_t$ to achieve accuracy of the mean Reynolds shear stress. That is, we can presume that the relaxation of the linear eddy-viscosity assumption is essential for successful learning. In fact, it was reported that the Vreman-type model, which is slightly free from linearity, performs somewhat better than DSM in channel turbulence \citep{Park2006,You2007}. The correlation coefficient between $|\bar{S}|^3$ and $\varepsilon_{SGS}$ listed in Table \ref{t.cor} clearly confirms that most DRL models and SVM produce similar levels of correlation, while DSM produces a high correlation. In LES wall modeling, artificial correlation between the resolved and modeled variables could cause LOW mismatch \citep{Yang2017}.

\begin{figure}
	\centerline{\includegraphics[width=1.0\columnwidth]{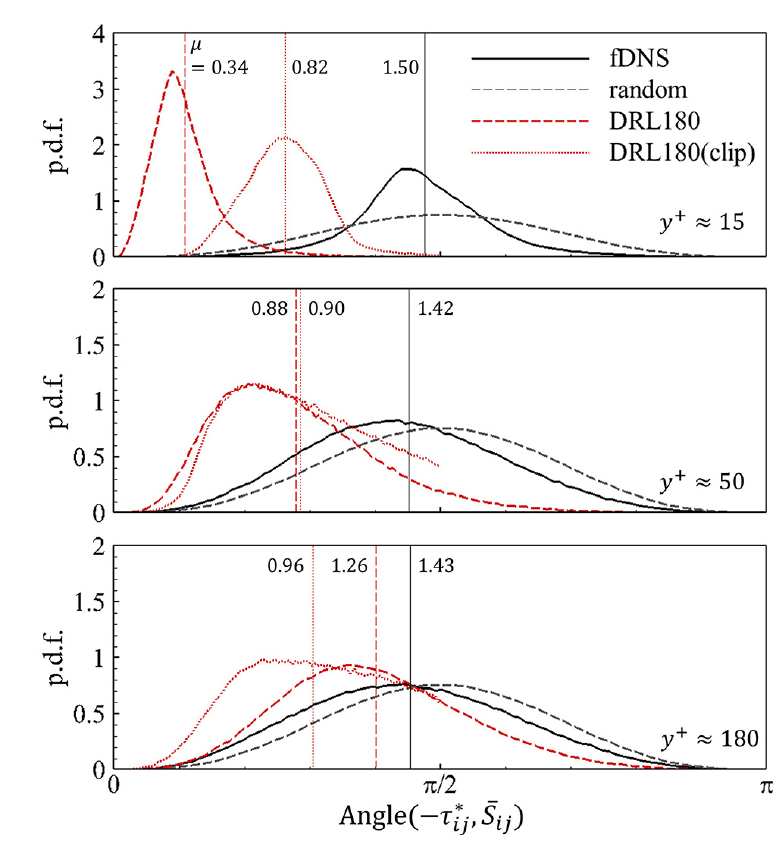}}
	\caption{PDFs of the angle between two tensors $\bar{S}_{ij}$ and $-\tau^*_{ij}$ at $y^+\approx 15, 50$, and $180$ in flow with $Re_\tau=180$ and $(\Delta x^+, \Delta z^+)=(70.7,35.3)$. The angle of the linear eddy-viscosity model is always zero.}
	\label{fig18}
\end{figure}

The major difference between the linear eddy-viscosity model including $\textrm{DRL180}(\nu_t)$ and other DRL SGS models is the alignment property between $\bar{S}_{ij}$ and $\tau_{ij}$; they are perfectly aligned in linear eddy-viscosity models, while there is no such restriction in other models. Whether such an alignment restriction is applied may critically affect the model prediction performance. Therefore, we investigate the angle distribution between $\bar{S}_{ij}$ and anisotropic parts of SGS stress tensor $\tau^*_{ij} (=\tau_{ij}-\tau_{kk}\delta_{ij}/3)$. The angle is defined as:
\begin{equation}
	\textrm{Angle}(-\tau^*_{ij}, \bar{S}_{ij}) = \textrm{arccos}\left({-\tau^*_{ij}\bar{S}_{ij} \over \sqrt{\tau^*_{kl}\tau^*_{kl}}\sqrt{\bar{S}_{mn}\bar{S}_{mn}}}\right) .
\end{equation}	
The PDFs according to the wall-normal location are shown in Fig. \ref{fig18}. Note that the angle of the eddy-viscosity model is always zero. Interestingly, the mean values of the angles in fDNS are near $\pi/2$ regardless of the wall-normal height. From a comparison with the PDF for two random tensors ($\sim \sin^3 \theta$ in a 5-dimensional space constructed by a trace-free symmetrical tensor), $\bar{S}_{ij}$ and $\tau^*_{ij}$ of fDNS tend to be orthogonal at $y^+ \approx 15$, but they are almost uncorrelated with each other in the region far from the wall, although the peak angle is slightly smaller than $\pi/2$. This implies that the supervised learning approach using only local strain-rate information as input might fail to reach successful local predictions, because the input and target are orthogonal or almost uncorrelated. On the other hand, the peak of the PDFs of the DRL180 model is found progressively from a value close to 0 (at 0.3 or $17^o$) toward $\pi/2$ with the wall-normal distance, whereas the mean value of the angle for DRL180(clip) is insensitively observed at approximately 0.9 ($\approx 50^o$) at all locations. This type of additional degree of freedom of DRL in the determination of $\tau_{ij}$ allows more flexible exploration in search of the optimal SGS model.

\begin{figure*}
	\centerline{\includegraphics[width=1.8\columnwidth]{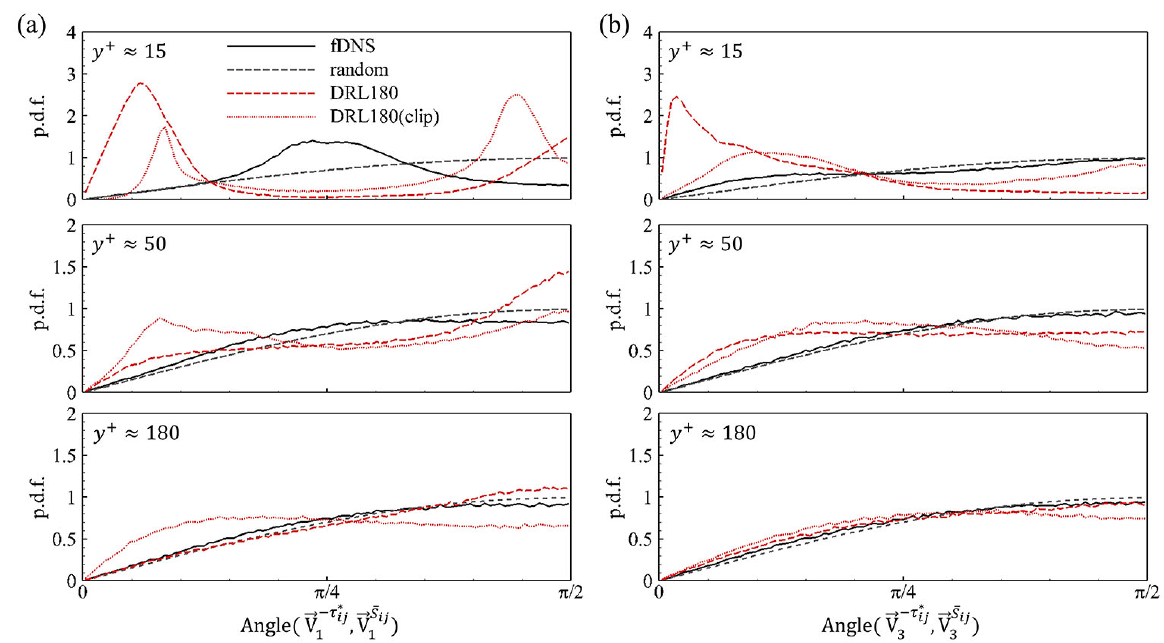}}
	\caption{PDFs of the angle between the eigenvectors of $\bar{S}_{ij}$ and $-\tau^*_{ij}$ at $y^+\approx 15, 50$, and $180$ in flow with $Re_\tau=180$ and $(\Delta x^+, \Delta z^+)=(70.7,35.3)$. The angle of the linear eddy-viscosity model is always zero. (a) $\vec{V}_1$, corresponding to the most stretching direction. (b) $\vec{V}_3$, corresponding to the most compressing direction.}
	\label{fig19}
\end{figure*}

A more detailed investigation of the alignment characteristics between $\bar{S}_{ij}$ and $\tau^*_{ij}$ can be performed by comparing the principal directions of both tensors. Thus, we compared the angle between the eigenvectors of $\bar{S}_{ij}$ and $-\tau^*_{ij}$, which are defined as
\begin{multline}
	\textrm{Angle}(\vec{V}_k^{-\tau^*_{ij}}, \vec{V}_k^{\bar{S}_{ij}}) = \\
	\textrm{arccos}\left({|\vec{V}_k^{-\tau^*_{ij}} \cdot \vec{V}_k^{\bar{S}_{ij}}| \over \sqrt{\vec{V}_k^{-\tau^*_{ij}} \cdot \vec{V}_k^{-\tau^*_{ij}}}\sqrt{\vec{V}_k^{\bar{S}_{ij}} \cdot \vec{V}_k^{\bar{S}_{ij}}}}\right),~~~~(no~sum~on~k)
\end{multline}
where $\vec{V}_k$ is the eigenvector of each tensor with corresponding eigenvalues $\lambda_k$ ($\lambda_1 \geq \lambda_2 \geq \lambda_3$). The PDFs of the angle between the most stretching eigenvectors of the two tensors ($\textrm{Angle}(\vec{V}_1^{-\tau^*_{ij}}, \vec{V}_1^{\bar{S}_{ij}})$) and the angle between the most compressive eigenvectors ($\textrm{Angle}(\vec{V}_3^{-\tau^*_{ij}}, \vec{V}_3^{\bar{S}_{ij}})$) are presented in Fig. \ref{fig19}. Near the wall, the angle in the most stretching direction of fDNS shows a peak at approximately $\pi/4$, while the angle in the most compressing direction of fDNS is almost randomly distributed. As the wall-normal height increased, the angle distribution of the fDNS in both directions approached a random one. However, the angle of the DRL model in both directions shows conspicuous behavior at $y^+ \approx 15$; both have a strong peak near 0, indicating that the most stretching and compressive eigenvectors of $-\tau^*_{ij}$ tend to align with the corresponding eigenvectors of $\bar{S}_{ij}$, but not perfectly. Since $\bar{S}_{xy}$ and $\bar{S}_{zy}$ near the wall are the two most dominant components among $\bar{S}_{ij}$, this alignment seems to be reasonable. However, as the wall-normal distance increases, such alignment behavior quickly disappears. Therefore, the perfect alignment assumption of a linear eddy-viscosity SGS model is limited. Indeed, the DRL model based on the eddy-viscosity assumption, $\textrm{DRL180}(\nu_t)$, did not perform well, as discussed previously. This confirms that relaxation of perfect alignment is essential for successful learning.

\begin{figure*}
	\centerline{\includegraphics[width=1.8\columnwidth]{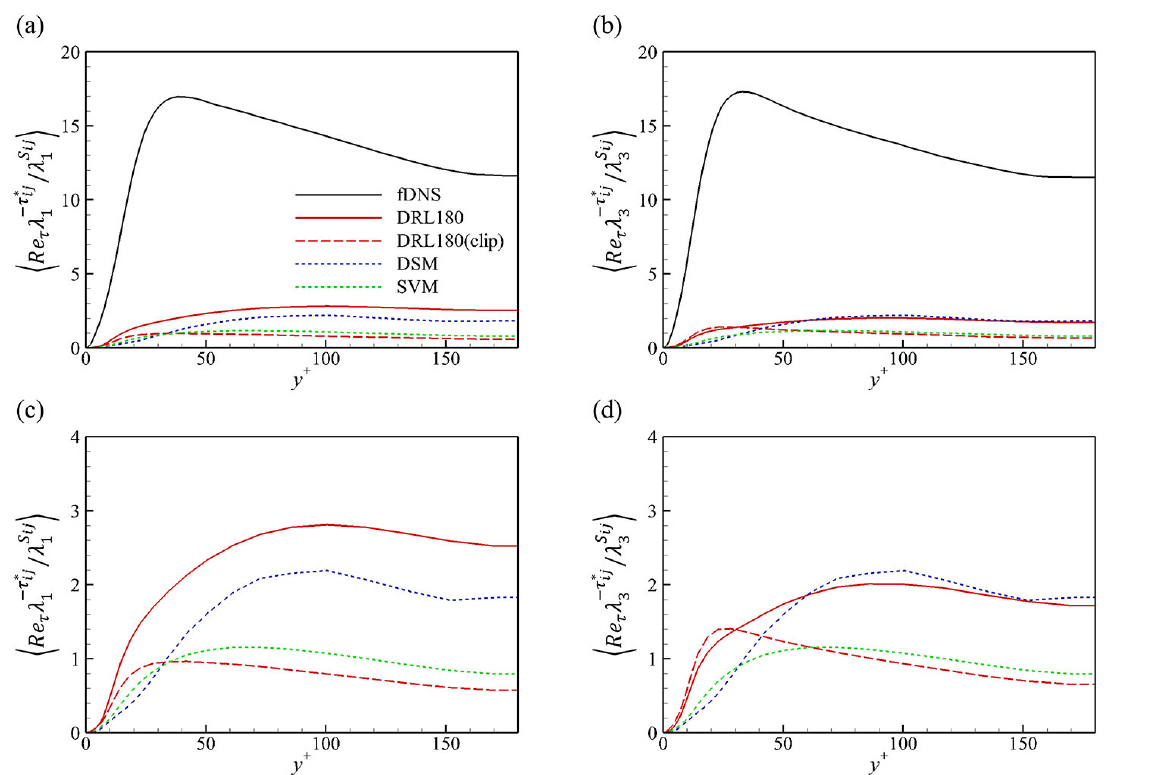}}
	\caption{Mean ratio between eigenvalues of $\bar{S}_{ij}$ and $-\tau^*_{ij}$ in flow with $Re_\tau=180$ and $(\Delta x^+, \Delta z^+)=(70.7,35.3)$ (a) in the most stretching direction, (b) in the most compressing direction, (c) a blowup of (a) for better comparison, and (d) a blowup of (b).}
	\label{fig20}
\end{figure*}

An additional investigation was carried out using the ratio of the eigenvalues of $-\tau^*_{ij}$ and $\bar{S}_{ij}$, as shown in Fig. \ref{fig20}, which has the physical meaning of twice the eddy viscosity averaged over the horizontal plane. The ratios in both the most stretching and compressive directions for fDNS, which are comparable to each other, are one order of magnitude larger than the predictions of all the models considered in this study, as shown in Fig. \ref{fig20}(a) and (b). Between the tested models, the level or distribution of the ratio is similar across all models, as shown in Fig. \ref{fig20}(c) and (d). It is interesting to note that the ratio for the DRL180 model and DSM has a similar distribution and that by DRL180(clip) and SVM shows a similar distribution in the region $y^+\geq 50$, while the predictions by DRL180 and DRL180(clip) are approximately twice as large as those by DSM and SVM in the region $y^+ \leq 15$. This analysis, combined with the alignment investigation, strongly suggests that the non-alignment nature between $-\tau^*_{ij}$ and $\bar{S}_{ij}$, rather than the level of eddy viscosity, plays a more important role in improving the performance of the SGS model. 

\section{Conclusions}\label{sec4}

In this study, we successfully developed DNN-based SGS models for the LES of turbulent channel flow through the DRL algorithm with physical constraints. The objective of DRL is to find the SGS model that maximizes the statistical accuracy of LES, and the mean of the viscous stress and Reynolds shear stress were used as target statistics. The reward function was defined as the negative distance between the weighted time-averaged statistics of the LES and fDNS statistics. The local velocity gradient tensor and grid size are the local state, and the statistics averaged over the homogeneous directions are the global state. The action consists of six components of SGS stress or a scalar eddy viscosity produced based on the local state. The training and testing results of the DRL are summarized as follows.

The TD3 algorithm with a one-step temporal difference reward is unstable in the learning of turbulent flows. The reward calculated using temporally accumulated statistics for DRL steps much larger than 1 with the discount factor $\gamma=0.99$ makes the learning stable and effective. This indicates that the reaction to the SGS stress is highly delayed, and the reward should include the long-term reaction more directly. Furthermore, we tested the proposed exploration method, the noise of which is spatially and temporally correlated, which showed the effectiveness compared with the cases without noise or with decorrelated Gaussian noise. In addition, imposing physical constraints on the DNN, such as reflection invariance and boundary conditions, makes exploration efficient and the possible solution space narrow. We confirmed that the proposed algorithm can discover the optimal SGS model in diverse LES environments.

Actual LESs with the trained SGS models were performed for statistical comparison with conventional SGS models. In all training environments, $\textrm{DRL180}$ and $\textrm{DRL180(clip)}$ were superior to the conventional models and $\textrm{DRL180}(\nu_t)$. $\textrm{DRL180}$ and $\textrm{DRL180(clip)}$ accurately predicted the target statistics, mean velocity, and mean Reynolds shear stress, while SVM and DSM overpredicted the magnitude of the mean Reynolds shear stress because of insufficient SGS dissipation, and $\textrm{DRL180}(\nu_t)$ overpredicted the mean velocity. This means that the eddy-viscosity assumption could seriously limit the performance, and backscatter is not necessary for predicting the given target statistics. Furthermore, the DRL models generated vortical structures similar to those of fDNS and provided better prediction accuracy of other Reynolds stress statistics and energy spectra than the conventional models. Importantly, we observed that mean SGS transport is predicted very accurately in the DRL model that provides reliable mean velocity and Reynolds shear stress, whereas mean SGS dissipation is not uniquely deterministic. That is, the mean SGS dissipation or backscatter might be related to various statistics and physics.

We tested the DRL model for flows of untrained Reynolds numbers and grid resolution. Overall, both $\textrm{DRL180}$ and $\textrm{DRL360}$ predicted the statistics of untrained Reynolds number flows with reasonable accuracy, whereas the DSM and 0-model highly overestimated the magnitude of the mean Reynolds stress for all Reynolds number flows. However, $\textrm{DRL180}$ trained at $Re_\tau=180$ slightly underestimated the mean velocity at a high Reynolds number, and $\textrm{DRL360}$ trained at $Re_\tau=360$ overpredicted it at $Re_\tau=180$. With regard to the generalization of the grid resolution, $\textrm{DRL180}$ trained in a single resolution environment did not predict the statistics of the flow with (coarse) untrained resolution well. The effect of grid resolution is very difficult to represent, as shown in the results of conventional SGS models. These results suggest that learning in multiple environments with different simulation parameters is necessary for developing a generally applicable and highly reliable model, which can be future work.

Through the analysis of the SGS stress produced by the trained DRL models, we determined its characteristics. One of the prominent features is that the correlation between the rate of strain tensor and the SGS stress tensor of DRL models near the wall is much lower than that of DSM, and a similar feature was also observed in the filtered DNS data. In other words, it is difficult to express the model obtained through DRL in a linear eddy-viscosity form because of the high variance in eddy viscosity. More importantly, in the PDFs of the angle between the strain-rate tensor and (negative) SGS stress tensor, a significant difference was observed between the linear eddy-viscosity model and the trained DRL model. The angle of the trained DRL model is quite far from zero and tends to become orthogonal as the wall distance increases. The alignment property between the filtered strain rate tensor and the SGS stress tensor seems to play a more important role than the level of the eddy viscosity in successful learning. This implies that to achieve reliable accuracy in DRL, we need to use a model-free form \citep{Gamahara2017}, general eddy-viscosity form \citep{Pope1975,Ling2016}, or mixed form. 

Several important issues must be considered in future studies. First, although deep learning is applied to various problems in fluid mechanics, the detailed principles of learning and the characteristics of trained neural networks are still unclear. In particular, reinforcement learning finds the optimal action (although not the global optimum) but does not provide an interpretable reason for how the action can be optimal. A deep understanding of deep learning is a well-known challenge because of the excessive number of parameters in neural networks; however, it is very important. This understanding can provide a guide to the tuning of hyperparameters involved in the design of neural networks, which is often very time-consuming. It has also been reported that the characteristics of the trained model may be closely related to the physics inherent in the data \citep{Kim2020b, Lu2020, Portwood2021}. This may be a problem that must be resolved for robustness and generalization of the model.

Moreover, deep learning-based SGS models were developed mostly in canonical flows, because most supervised learning requires DNS data for training. There is no guarantee that the model trained in a canonical flow works well in other flows with different physical characteristics; thus, the learning of various flows where DNS flow fields are usually not available is necessary for generalization. As shown in the present study of wall-bounded turbulence, DRL that requires only statistical information in the training process would be an alternative algorithm for developing effective SGS models in practical applications. However, an extension of the DRL to complex flows is challenging in many respects. First, because classical CNNs without interpolation are inappropriate in irregular domains, special types of DNNs such as the graph CNN \citep{Kipf2016} and PointNet \citep{Qi2017,Kashefi2021} would be needed. Second, in channel turbulence, a straightforward application to different Reynolds number flows based on wall unit scaling is possible, but in complex flow, the characteristic velocity and length scales change between problems, and the trained NN-based SGS model might not work with the change of scaling. Furthermore, our model did not implement the rotational invariance, which might be essential for generalizability of the developed model. As a good example, the local framework that the NN-based SGS model satisfies the unit, reflectional, and rotational invariances was proposed \citep{Prakash2021}. Another important issue is that target statistics in complex flows might be partially available (in space and time) or noise-added because of measurement difficulties. An investigation into whether successful DRL is possible even with an incomplete target or how much target information is required would be interesting future work. Finally, for the development of an SGS model capable of successful prediction in various complex flows, training of the model and running of diverse flow solvers should be conducted simultaneously. A platform that includes the construction of diverse flow solvers, its parallelization, and combination with the DRL algorithm needs to be developed. 

Recently, another approach called a differentiable PDE method \citep{Sirignano2020, Kochkov2021,Stroefer2021,List2022} has emerged as a promising algorithm for turbulence modeling. Although this method has disadvantages compared with DRL in that it needs to build a differentiable PDE solver and to memorize all three-dimensional flow data, it is an interesting technique that guarantees robust performance, being superior to a standard supervised learning. Therefore, a quantitative comparison of the DRL and the differentiable PDE method would be worthwhile from various perspectives.

We clearly demonstrated that the DRL framework is a promising algorithm for developing an SGS model of LES. With an increase in available computational resources, the performance of the model will naturally improve. Although we have shown successful applicability only in channel flow, this framework can be extended to various turbulence modeling problems such as the SGS modeling of LES in complex geometries, scalar flux modeling \citep{Frezat2021}, and LES wall modeling \citep{Yang2019}. The ultimate goal of DRL-based LES modeling is to search for a universal SGS model that can be applied to various problems with reasonable accuracy and stability through accumulated learning experiences.

\vspace{0.3in}
\noindent
{\bf {ACKNOWLEDGMENTS}}

This work was supported by a National Research Foundation of Korea (NRF) grant funded by the Korean government (MSIP) (2017R1E1A1A03070282, 2022R1A2C2005538).

\vspace{0.3in}
\noindent
{\bf AUTHOR DECLARATIONS} \\
{\bf Conflict of Interest} 

\noindent
The authors have no conflicts to disclose. 

\vspace{0.1in}
\noindent
{\bf Author Contributions} 

\noindent
Junhyuk Kim: Conceptualization (equal); Formal analysis (lead); Investigation (equal); Methodology (equal); Computing (lead); Writing - original draft (lead); Writing - review \& editing (equal). Hyojin Kim: Formal analysis (equal); Investigation (equal). Jiyeon Kim: Formal analysis (equal); Investigation (equal). Changhoon Lee: Conceptualization (equal); Formal analysis (supporting); Funding acquisition (lead); Investigation (equal); Methodology (equal); Project administration (lead); Software (lead); Supervision (lead); Writing - review \& editing (lead).

\vspace{0.1in}
\noindent
{\bf {DATA AVAILABILITY}} \\
\noindent
The data that support the findings of this study are available from the corresponding author upon reasonable request.

\vspace{0.3in}
\noindent
{\bf REFERENCES}
\bibliography{ref}

\end{document}